\documentclass[a4paper, 11pt]{article}
\usepackage{
  jheppub
} 
\usepackage{lineno}
\usepackage{booktabs} 
\usepackage{graphicx}
\usepackage{float}
\usepackage{array}
\usepackage{extarrows}
\usepackage{amssymb}
\usepackage{amsmath}
\usepackage{subcaption}
\usepackage{makecell}
\usepackage{multirow}
\usepackage{standalone}
\usepackage[compat=1.1.0]{tikz-feynman}

\arxivnumber{2607.03249} 

\title{\boldmath Probing a 146 GeV cLFV scalar using the LHC and low-energy experiments}

\author[1]{Christina Gao,}
\author[2,3]{Lingfeng Li,}
\author[1]{Z.J. Xiong}
\affiliation[1]{Department of Physics, Southern University of Science and Technology, Shenzhen, Guangdong 518055, China}
\affiliation[2]{Nanjing Normal University, Nanjing, Jiangsu, 210023, China}
\affiliation[3]{International Center for Theoretical Physics Asia-Pacific, Beijing, 100190, China}

\emailAdd{gaoy3@sustech.edu.cn, l.f.li165@gmail.com, 12432073@mail.sustech.edu.cn}

\abstract{%
The CMS Collaboration reported a local excess at $146~\mathrm{GeV}$ in the search for the lepton-flavor-violating decay of the Higgs boson and additional Higgs bosons in the $e\mu$ final state at $\sqrt{s}= 13~\mathrm{TeV}$. If confirmed, this would constitute a major piece of evidence of charged lepton flavor violation (cLFV). We investigate the compatibility of the claimed signal with the full suite of existing low-energy cLFV constraints in a bottom-up effective description: a single real scalar of mass $146~\mathrm{GeV}$ coupled to gluons and to all charged-lepton bilinears, with seven free parameters that simultaneously control the LHC production cross section, every di-lepton decay channel, and every low-energy cLFV observable. A Bayesian MCMC analysis against $\mu-e$ conversion, muonium-antimuonium oscillation, three-lepton and radiative LFV decays, semileptonic $\tau$ LFV decays, and LHC di-lepton searches yields a preferred mode with peaked value $Y_{e\mu} \sim 10^{-4.09}$, already cut into by the current $\mu-e$ conversion limits. The projected sensitivities of Mu2e, COMET, Mu3e, MACE, MEG~II, Belle~II, STCF, and the HL-LHC directly probe the region of coupling space selected by the CMS excess, so the complementarity between high-energy and low-energy cLFV probes will either corroborate or decisively exclude the scalar interpretation of the anomaly within the next decade. }

\begin{document}
  \maketitle
  \newpage
  \flushbottom

  \section{Introduction}
  \label{sec:intro}

  The CMS Collaboration has reported a local excess in the search for the lepton-flavor-violating decay of the Higgs boson and additional Higgs bosons in the $e\mu$ final state, using $138~\mathrm{fb}^{-1}$ of proton-proton collision data at $\sqrt{s}= 13~\mathrm{TeV}$~\cite{CMS:2023pte}. The excess is consistent with a new scalar resonance of mass $146~\mathrm{GeV}$ decaying into $e\mu$, with a global (local) significance of $2.8\sigma$ ($3.8\sigma$) and an inferred cross section $\sigma(pp \to \phi \to e\mu) \simeq 3.89~\mathrm{fb}(6~\mathrm{fb})$. No corresponding excess is observed in the Standard Model (SM) Higgs boson decay $h \to e\mu$, and the ATLAS Collaboration has not reported an analogous feature. If confirmed by future data, this would constitute a major piece of evidence of charged lepton flavor violation (cLFV) in any experiment.

  The SM contains no source of cLFV: contributions induced by neutrino masses are suppressed by the GIM mechanism to branching ratios of $\mathcal{O}(10^{-54})$~\cite{Marciano:1977wx,Petcov:1976ff,Bilenky:1977du,Cheng:1977nv,Lee:1977qz}, rendering any signal unambiguously beyond the SM. Decades of dedicated low-energy searches have returned null results and placed stringent limits on the same operators that would produce a $146~\mathrm{GeV}$ resonance decaying to $e\mu$. These include searches for $\mu \to e\gamma$ at MEG~II~\cite{MEGII:2023ltw}, $\mu \to eee$ at SINDRUM~\cite{SINDRUM:1987nra} with the substantial improvement expected from Mu3e~\cite{Hesketh:2022wgw}, $\mu-e$ conversion in nuclei at SINDRUM~II~\cite{SINDRUMII:2006dvw} with projected gains at Mu2e~\cite{Miscetti:2025uxk} and COMET~\cite{Moritsu:2022lem}, muonium-antimuonium oscillation at MACS~\cite{Willmann:1998gd} with a next-generation program at MACE~\cite{Bai:2024skk}, and $\tau$ LFV decays at Belle~\cite{Belle:2021ysv}, Belle~II~\cite{Belle-II:2025urb, Belle-II:2024sce}, LHCb~\cite{LHCb:2026eod}, ATLAS~\cite{ATLAS:2026kkw}, and the proposed STCF~\cite{Achasov:2023gey}. With a mediator near the electroweak scale, the mass scale of new physics alone does not suppress any of these processes, so reconciling the CMS hint with the null results requires a specific arrangement of couplings. Latest global constraints for those operators have been performed in \cite{Delzanno:2024ooj,Fernandez-Martinez:2024bxg, Abu-Ajamieh:2025jsz}. The sensitivity of future colliders to those operators has also been studied~\cite{Dev:2017ftk,Li:2018cod,Xu:2023ene,Arroyo-Urena:2025dxa, Arroyo-Urena:2025mxv}. Theoretically, such cLFV operators can arise in a variety of BSM scenarios. Recent works have investigated those including 2HDM~\cite{Arroyo-Urena:2023vfh}, leptoquark~\cite{Varzielas:2023qlb}, charged Higgs~\cite{Arroyo-Urena:2024soo}, vector boson~\cite{Huang:2025dga}, and Flavon~\cite{Koivunen:2023sie, Han:2025jqx, Arroyo-Urena:2025gcs} models.  For more discussions on general cLFV, see Refs.~\cite{Calibbi:2017uvl,Ardu:2022sbt,Frau:2024rzt}.

  Several explanations of the $146~\mathrm{GeV}$ hint have been proposed within specific ultraviolet frameworks. Koivunen and Raidal~\cite{Koivunen:2023sie} have studied the Froggatt--Nielsen mechanism and shown that the purely leptophilic flavon cannot reach the reported cross section, while a version with additional flavon-quark couplings requires a tuned cancellation between top and charm contributions to $\mu-e$ conversion. Primulando et al.~\cite{Primulando:2023ohe} have addressed the same signal within type-III two-Higgs-doublet models, and Afik, Dev, and Thapa~\cite{Afik:2023vit} have proposed a leptophilic two-Higgs-doublet scenario in which the new scalars are produced resonantly through the lepton content of the proton rather than through gluon-gluon fusion (ggF). Liu and Ivanov \cite{Liu:2026voi} have embedded the signal in a CP4-symmetric three-Higgs-doublet model, and identified a viable scenario that accommodates the $146~\mathrm{GeV}$ excess while satisfying current low-energy cLFV constraints. Each of these studies is valuable, but they commit to a particular UV framework, or to a particular production mechanism, or to the assumption of SM-Higgs-like couplings, making the question of whether the CMS signal is compatible with the full suite of cLFV bounds conditional on those choices. A complementary and logically prior question is whether \emph{any} scalar mediator, independent of its UV origin, can accommodate the signal through ggF production, and which regions of its coupling space survive the existing constraints. We address this question here in a bottom-up effective description. We introduce a single real scalar $\phi$ of mass $M_{\phi}= 146~\mathrm{GeV}$ with the Lagrangian
  \begin{equation}
    \mathcal{L}_{\rm LFV}\supset - \kappa_{gg}\, \phi\, G^{a}_{\mu\nu}G^{a\,\mu\nu}
    - \sum_{i}Y_{ii}\, \phi\, \bar\ell_{i}\ell_{i}- \sum_{i \ne j}Y_{ij}\, \phi\,
    \bar\ell_{i}\ell_{j}, \label{eq:Lintro}
  \end{equation}
  with $i,j \in \{e,\mu,\tau\}$ and $Y_{ij}= Y_{ji}$ assumed real. The seven parameters $\kappa_{gg}$, $Y_{ee}$, $Y_{\mu\mu}$, $Y_{\tau\tau}$, $Y_{e\mu}$, $Y_{e\tau}$, $Y_{\mu\tau}$ simultaneously control the ggF production cross section at the LHC, every di-lepton decay channel of $\phi$, and low-energy cLFV observables, so the full experimental suite can be imposed as a single consistency test without commitment to a UV completion.

  The seven-parameter structure allows us to make the complementarity between high-energy and low-energy probes quantitative: the cross section measurement at the CMS $146~\mathrm{GeV}$ excess fixes a specific combination of $\kappa_{gg}$, $Y_{e\mu}$ and the total width $\Gamma_{\phi}$, whereas the remaining directions in coupling space are constrained by experiments of an entirely different character. Performing a Bayesian scan of the seven-parameter space against all of these observables simultaneously, we find that the projected sensitivities of the next generation of low-energy experiments, such as $\mu-e$ conversion at Mu2e~\cite{Miscetti:2025uxk} and COMET~\cite{Moritsu:2022lem}, together with $\mu \to eee$ at Mu3e~\cite{Hesketh:2022wgw}, directly probe the region of $Y_{e\mu}$ hinted by the CMS signal. The implication is that in the future, data from $e^+e^-$ colliders, muon experiments and HL-LHC will either pin down the $146~\mathrm{GeV}$ anomaly or rule it out, regardless of which UV completion is ultimately responsible.

  The remainder of this paper is organized as follows. In Sec.~\ref{sec:model} we define the toy model and fix notation. In Sec.~\ref{sec:observables} we derive the contributions of the model to the LHC ggF cross section and to the representative cLFV processes entering the global analysis: $\mu-e$ conversion, muonium-antimuonium oscillation, three-lepton LFV decays, radiative LFV decays, and semileptonic $\tau$ LFV decays. In Sec.~\ref{sec:analysis} we describe the Bayesian MCMC setup, the likelihood construction, and the numerical results. We conclude in Sec.~\ref{sec:conclusion}.

  \section{Toy Model}
  \label{sec:model}
  We collect here the definitions and assumptions underlying the effective Lagrangian introduced in Eq.~\eqref{eq:Lintro}, and briefly indicate how it arises as a low-energy limit of several ultraviolet completions proposed in the literature.

  The field content is a single real scalar $\phi$ of mass $M_{\phi}= 146~\mathrm{GeV}$, coupled to gluons through an effective dimension-five operator and to the SM charged leptons through flavor-general Yukawa interactions,
  \begin{equation}
    \mathcal{L}_{\rm LFV}\supset - \kappa_{gg}\, \phi\, G^{a}_{\mu\nu}G^{a\,\mu\nu}
    - Y_{ee}\, \phi\, \bar{e}e - Y_{\mu\mu}\, \phi\, \bar{\mu}\mu - Y_{\tau\tau}\,
    \phi\, \bar{\tau}\tau - (Y_{e\mu}\, \phi\, \bar{e}\mu + Y_{e\tau}\, \phi\, \bar
    {e}\tau + Y_{\mu\tau}\, \phi\, \bar{\mu}\tau + \textrm{h.c.}), 
  \label{eq:Ltoy}
  \end{equation}
  where $G^{a}_{\mu\nu}$ is the gluon field strength. The gluon coupling $\kappa_{gg}$ has mass dimension $-1$ and arises physically from integrating out heavy colored states \cite{Dawson:2014ora,Cai:2018cog}; the Yukawa couplings $Y_{ij}$ are dimensionless.\footnote{Strictly speaking, $Y_{ij}$ shall come from the dimension-5 couplings of the form $\phi (\bar{L}_i H \ell_j+\bar{L}_j H \ell_i)/\Lambda$, where $L$($H$) is the SM left-handed lepton (Higgs) doublet and $\Lambda$ is the cutoff scale.} The lepton bilinears are understood to be written in the physical mass basis. The potential of $\phi$ is chosen such that it does not acquire a vacuum expectation value.

  We work throughout under the following simplifying assumptions. All seven couplings $\{\kappa_{gg}, Y_{ee}, Y_{\mu\mu}, Y_{\tau\tau}, Y_{e\mu}, Y_{e\tau}, Y_{\mu\tau} \}$ are taken to be real, so $\phi$ is CP-even and no $\phi\, G \widetilde{G}$ operator is generated; the off-diagonal Yukawas are symmetric, $Y_{ij}= Y_{ji}$; and $\phi$ is treated as an on-shell resonance in the narrow-width approximation throughout. Assigning $\kappa_{gg}$ and the $Y_{ij}$ as free parameters rather than deriving them from a specific UV sector is the price of the model-agnostic posture: our analysis constrains the operators \emph{as written}, and the mapping to parameters of any particular UV completion lies outside the scope of this work.

  The effective structure of Eq.~\eqref{eq:Ltoy} is reproduced, in an appropriate limit, by several UV frameworks that have been invoked to address the CMS $146~\mathrm{GeV}$ hint. In the Froggatt--Nielsen construction of Ref.~\cite{Koivunen:2023sie}, the flavon is a complex scalar whose real part acquires off-diagonal lepton couplings through the messenger-induced Yukawa structure; adding flavon-quark couplings induces an effective $\phi\, G G$ operator through top loops, reproducing $\kappa_{gg}$ in our notation. In the type-III two-Higgs-doublet model of Ref.~\cite{Primulando:2023ohe}, the heavy neutral scalar plays the role of $\phi$, with the off-diagonal Yukawas arising from the second-doublet Yukawa matrix and $\kappa_{gg}$ generated by a top loop proportional to the doublet's quark couplings.
  A global fit of flavor-violating charged-lepton Yukawa couplings has been performed in Ref.~\cite{Abu-Ajamieh:2025jsz} under the assumption of SM-Higgs-like production. In what follows, we treat Eq.~\eqref{eq:Ltoy} directly and report bounds on its seven parameters.

  \section{Observables}
  \label{sec:observables}
  The seven parameters of Eq.~\eqref{eq:Ltoy} enter two classes of observables. The first is the LHC production and decay of $\phi$ itself, which for a $146~\mathrm{GeV}$ scalar produced in ggF is controlled by the combination $\kappa_{gg}^2\times\textrm{Br}(\phi\to X)$ and is therefore sensitive to every Yukawa coupling through the total width. The second is a set of low-energy processes in which $\phi$ appears only as a virtual mediator, contributing at tree level to $\mu-e$ conversion, three-lepton LFV decays, muonium-antimuonium oscillation, and semileptonic $\tau$ LFV decays, and at one loop to radiative LFV decays. Each of these observables depends on a specific combination of $\kappa_{gg}$ and $Y_{ij}$, so the $146~\mathrm{GeV}$ signal and the low-energy bounds carve the parameter space along different directions.

  The calculations for the cross section or width of the relevant observables are well understood. Here we review each of them in turn.

  \subsection{Production of $\phi$ at the LHC}

  The relevant production mechanism for $\phi$ at the LHC is ggF through the effective $\phi GG$ operator. The corresponding cross section is the convolution of the gluon parton distribution functions (PDFs) $f_{g}$ with the parton-level cross section $\hat{\sigma}_{gg \to \phi}$:

  \begin{equation}
    \sigma_{pp \to \phi}(s)= \int_{0}^{1}\mathrm{d}x_{1}\int_{0}^{1}\mathrm{d}x_2\ f_{g}(x_{1},\mu_{F}) f_{g}(x_2,\mu_{F}) \hat{\sigma}_{gg \to \phi}(\hat{s}=x_{1}x_2s)\,,
  \end{equation}
  where $\mu_{F}=M_{\phi}$ is the factorization scale and $s=(13 \textrm{ TeV})^2$ is the center-of-mass energy squared of the LHC search~\cite{CMS:2023pte}. The leading order (LO) parton-level cross section $\hat{\sigma}_{gg \to \phi}(\hat{s})$ can be expressed as:
  \begin{equation}
    \hat{\sigma}_{gg \to \phi}^{\textrm{LO}}(\hat{s}) = \frac{\pi^2}{8 M_{\phi}}\delta(\hat{s}-M_{\phi}^2) \Gamma_{\phi\to gg}^{\textrm{LO}}\,.
  \end{equation}
  Then the proton-level LO cross section becomes:
  \begin{align}
    \sigma_{pp \to \phi}^{\textrm{LO}}(s,M_{\phi}^2) & = \int_{0}^{1}\mathrm{d}x_{1}\int_{0}^{1}\mathrm{d}x_2\ f_{g}(x_{1},\mu_{F}) f_{g}(x_2,\mu_{F}) \frac{\pi^2}{8M_{\phi}}\delta(x_{1}x_2s-M_{\phi}^2)\Gamma_{\phi\to gg}^{\textrm{LO}}\notag \\
     & =\int_{\tau_\phi}^{1}\frac{\mathrm{d}x_{1}}{x_{1}}\ f_{g}(x_{1},\mu_{F}) f_{g}(\tau_{\phi}/x_{1},\mu_{F}) \frac{\tau_{\phi}\pi^2}{8 M_{\phi}^{3}}\Gamma_{\phi\to gg}^{\textrm{LO}}\notag           \\
    & = I_{gg}(s,M_{\phi}^2)\frac{\pi^2}{8 M_{\phi}^{3}}\Gamma_{\phi\to gg}^{\textrm{LO}}\,,
  \end{align}
  where $\tau_{\phi}=M_{\phi}^2/s$ and the dimensionless integral $I_{gg}$ is:
  \begin{equation}
    I_{gg}(s,M_{\phi}^2)=\int_{\tau_\phi}^{1}\frac{\mathrm{d}x_{1}}{x_{1}}\ f_{g}(x_{1},\mu_{F}) f_{g}(\tau_{\phi}/x_{1},\mu_{F})\tau_{\phi}\,.
  \end{equation}
  $I_{gg}$ can be computed using a numerical PDF evaluation tool ManeParse~\cite{Clark:2016jgm}.
  The leading order decay width of $\phi\to gg$ for this model \eqref{eq:Ltoy} is
  given by:
  \begin{equation}
    \Gamma_{\phi\to gg}^{\textrm{LO}}= \frac{2\kappa_{gg}^2 M_{\phi}^3}{\pi}\,.
  \end{equation}
  Thus the proton-level LO cross section can be expressed as:
  \begin{equation}
    \sigma_{pp \to \phi}^{\textrm{LO}}  = \frac{\pi\kappa_{gg}^2}{4}I_{gg}(s,M_{\phi}^2)\,.
  \end{equation}

  To include higher-order corrections, we introduce a $K$-factor following Ref.~\cite{Herzog:2017dtz}:
  \begin{align}
    K= & 1+(5.703052-1.220188 L_{\phi}) \alpha_{s}(M_{\phi}^2)\notag                                            \\
       & +(15.887961-0.578375 L_{t\phi}-10.927911L_{\phi}+1.116644L_{\phi}^2)\alpha_{s}^2(M_{\phi}^2)\notag \\
    =  & 1.47153\textrm{,}
  \end{align}
  where $L_{\phi}=\ln(M_{\phi}^2/\mu^2)$, $L_{t\phi}=\ln(m_{t}^2/M_{\phi}^2)$ and $\mu=M_{\phi}/2$. We set $\alpha_{s} (M_{\phi}^2)\approx 0.1095$ from PDG 2024 \cite{ParticleDataGroup:2024cfk}. Strictly speaking, the coefficients in the $K$-factor above are computed for SM Higgs production in the heavy-top effective theory ($m_t\to\infty$), where the $hGG$ vertex originates from integrating out the top quark loop~\cite{Herzog:2017dtz}. In our framework the $\phi GG$ operator is instead an independent dimension-five coupling $\kappa_{gg}$ whose UV origin is left unspecified. Nevertheless, the QCD radiative corrections to the $gg\to\phi$ process are governed by the same gluonic operator and are therefore identical at the level of the heavy-top EFT; the $K$-factor inherits a residual model dependence only through the finite-$m_t$ corrections, which for $M_\phi=146~\mathrm{GeV}\ll 2m_t$ are at the percent level~\cite{Dawson:2014ora}. We therefore adopt the SM $K$-factor as a reliable approximation and assign the resulting theoretical uncertainty as subdominant compared to the experimental uncertainties.
  Then the total theoretical cross section of production for final state $X$ at the LHC through the ggF process is given in the narrow-width approximation by

  \begin{equation}
    \sigma_{pp \to \phi \to X}^{\textrm{th}}= K\sigma_{pp \to \phi}^{\textrm{LO}}\cdot \textrm{Br}_{\phi \to X}\,.
  \end{equation}

  \subsection{$\phi$-mediated Low Energy cLFV Processes}
  \label{sec:LFV}

  The low-energy observables fall into two sub-classes. Those mediated by a tree-level $\phi$ exchange, including $\mu-e$ conversion, three-lepton decays, muonium-antimuonium oscillation, and semileptonic $\tau$ decays, probe specific Yukawa products directly, so their experimental bounds translate into clean exclusions in the corresponding coupling plane. The one-loop radiative decays $\ell_{\alpha}\to\ell_{\beta}\gamma$ instead probe Yukawa products through a sum over intermediate leptons and therefore mix several couplings at once; the resulting constraint turns out to be the tightest in the $(Y_{e\tau},Y_{\mu\tau})$ plane, as we will see in Sec.~\ref{sec:analysis}.

  \subsubsection{$\mu-e$ Conversion}

  \begin{figure}[htbp]
    \includegraphics[width=0.5\textwidth]{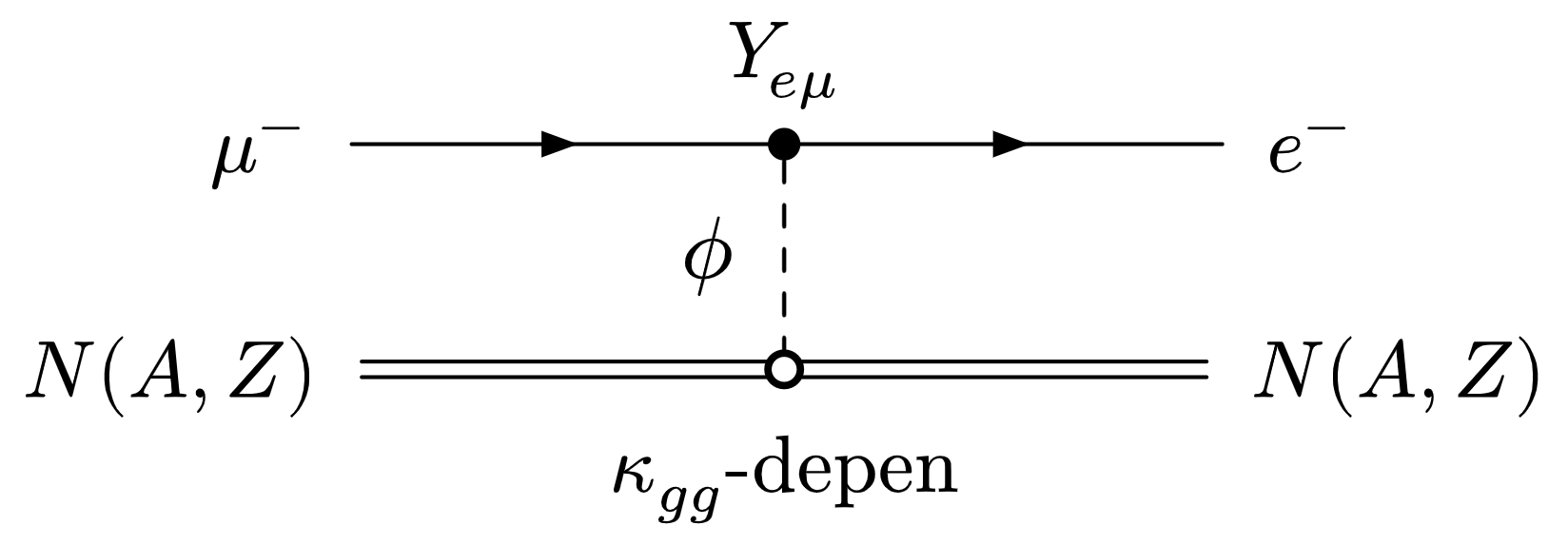}
    \centering
    \caption{Diagram for $\mu-e$ conversion via scalar exchange.}
    \label{fig:mueC}
  \end{figure}
  
  Among the low-energy observables, $\mu-e$ conversion is the only one that receives a tree-level contribution involving both $\kappa_{gg}$ and $Y_{e\mu}$ directly, as shown in Fig.~\ref{fig:mueC}, through the effective $\bar{e}\mu\,G G$ operator of Eq.~\eqref{eq:Leff-mue} below. It therefore links the LHC cross section of the $146~\mathrm{GeV}$ scalar to the strongest low-energy cLFV bound, and drives much of the complementarity visible in the global fit.

  \begin{equation}
    \label{eq:Leff-mue}\mathcal{L}_{\textrm{eff};\mu-e}= -\frac{\kappa_{gg}Y_{e\mu}}{M_{\phi}^2}\bar{e}\mu G_{a}^{\mu\nu}G^{a}_{\mu\nu}+\textrm{h.c.}\,.
  \end{equation}

  Rearrange the Lagrangian to match the conventions of Ref.~\cite{Cirigliano:2009bz}:
  \begin{equation}
    \mathcal{L}_{\textrm{eff};\mu-e}= -\frac{1}{M_{\phi}^2}C_Gm_{\mu}G_{F}\bar{e}\mu\frac{\beta_{L}}{2g_{s}^{3}}G_{a}^{\mu\nu}G^{a}_{\mu\nu}+\textrm{h.c.}\,,
  \end{equation}
  where $\beta_L = (g_{s}^{3}/16\pi^2)(11-2N_{f}/3)$ and $N_{f}$ counts light quarks only, and the coefficient $C_G$ reads
  \begin{equation}
    C_G= \frac{\kappa_{gg}Y_{e\mu}}{m_{\mu}G_{F}}\frac{2g_{s}^{3}}{\beta_{\textrm{L}}}\,.
  \end{equation}
  At the nucleon level $N = p,n$, the gluonic operator is replaced by a scalar density through
  \begin{equation}
    \frac{\beta_{L}}{2g_{s}^{3}}G_{a}^{\mu\nu}G^{a}_{\mu\nu}\longrightarrow f_{GN}m_{N}\bar{N}N\,,
  \end{equation}
  where the form factor
  \begin{equation}
    f_{GN}= 1 - \sum_{q=u,d,s}f^{q}_{SN}\label{eq:fGN-def}~,
  \end{equation}
  encodes the couplings of the gluonic operator to the light quarks in the nucleon through the scalar form factors $f^{q}_{SN}$. The evaluation of these form factors using lattice-QCD inputs from FLAG is collected in App.~\ref{app:fGN}; using the $N_{f}= 2+1$ FLAG 2024 averages~\cite{FlavourLatticeAveragingGroupFLAG:2024oxs} one obtains 
  \begin{equation}
    \label{eq:fGN-values}f_{Gp}\approx0.90869,\quad f_{Gn}\approx0.90565.
  \end{equation}
  The effective Lagrangian at the nucleon level is therefore
  \begin{equation}
    \mathcal{L}_{\textrm{eff};\mu-e}^{(N)}= -\sum_{N=p,n}\frac{m_{\mu}m_{N}}{M_{\phi}^2}f_{GN}C_G\bar{e}\mu \bar{N}N +\textrm{h.c.}\,.
  \end{equation}
  To compute the conversion rate, the relevant matrix elements are
  \begin{equation}
    \langle A,Z|\bar{p}p|A,Z \rangle=Z \rho^{(p)}, \quad \langle A,Z|\bar{n}n|A,Z\rangle= (A-Z)\rho^{(n)}\,,
  \end{equation}
  where $\rho^{(N)}$s are nucleon densities. The conversion rate is then given
  by~\cite{Cirigliano:2009bz}
  \begin{equation}
    \Gamma_{\mu-e\textrm{ conv}}= \frac{m_{\mu}^{5}}{2M_{\phi}^{4}}\left|4G_{F}m_{\mu}C_G\left(m_{p}f_{Gp}S^{(p)}+m_{n}f_{Gn}S^{(n)}\right)\right|^2\,,
  \end{equation}
  where $S^{(N)}$ is the overlap integral of muon and electron wave functions with nucleon density depending on $\rho^{(N)}$ and $A,Z$ \cite{Kitano:2002mt}; the most recent determinations of $S^{(N)}$ are given in Ref.~\cite{Borrel:2024ylg}. The branching ratio relative to muon capture is therefore
  \begin{equation}
    \textrm{Br}_{\mu-e\textrm{ conv}/(\textrm{A,Z})}^{\textrm{th}}= \frac{\Gamma_{\mu-e\textrm{ conv}/(\textrm{A,Z})}}{\Gamma_{\mu\textrm{ capt}/(\textrm{A,Z})}}\,.
  \end{equation}

  \subsubsection{Muonium-Antimuonium Oscillation}

\begin{figure}[htbp]
    \includegraphics[width=0.35\textwidth]{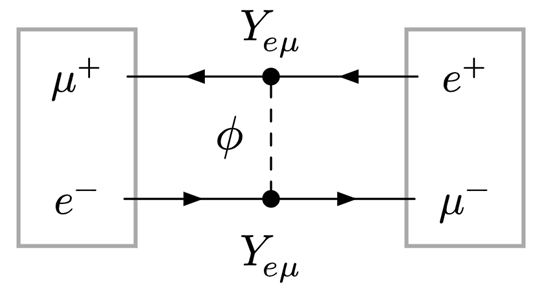}
    \centering
    \caption{Diagram for scalar mediated $M_\mu-\overline{M_\mu}$ oscillation.}
    \label{fig:M2M}
  \end{figure}

  Muonium ($M_\mu$), a hydrogen-like bound state of $\mu^+$ and $e^-$, and antimuonium ($\overline{M_\mu}$), the corresponding charge-conjugated system composed of $\mu^-$ and $e^+$, can oscillate into each other via a tree-level $\phi$ exchange between two $Y_{e\mu}$ couplings, as shown in Fig.~\ref{fig:M2M}. The oscillation probability depends on $Y_{e\mu}^4$, making it a sensitive probe of the same off-diagonal Yukawa that the LHC signal fixes.

  The time evolution of the $M_\mu-\overline{M_\mu}$ system is governed by a $2\times 2$ Hamiltonian matrix $\mathcal{M}$~\cite{Bai:2022sxq},
  \begin{equation}
    \mathrm{i}\frac{\mathrm{d}}{\mathrm{d}t}\left[\begin{matrix}|M_\mu(t)\rangle \\|\overline{M_\mu}(t)\rangle\end{matrix}\right]= \mathcal{M}\left[\begin{matrix}|M_\mu(t)\rangle \\|\overline{M_\mu}(t)\rangle\end{matrix}\right]\,.
  \end{equation}
  The diagonal terms of $\mathcal{M}$ are:
  \begin{equation}
    \mathcal{M}_{11}=\mathcal{M}_{22}=m_M-\mathrm{i}\Gamma_{M}/2\,.
  \end{equation}
  where $m_M,\Gamma_{M}$ are the mass and decay width of $M_\mu$ and one usually takes
  \begin{equation}
    m_M=m_{\mu}+m_{e}-m_{\textrm{bind}}\approx m_{\mu},\quad \Gamma_{M}\approx \Gamma_{\mu}\,.
  \end{equation}
  The off-diagonal terms are given by~\cite{Bai:2022sxq}:
  \begin{align}
    \mathcal{M}_{12}=\mathcal{M}_{12}^{*} & =\frac{1}{2m_M}\langle\overline{M_\mu}|\mathcal{H}_{\textrm{eff}}|M_\mu\rangle+\frac{1}{2m_M}\sum_{n}\frac{\langle\overline{M_\mu}|\mathcal{H}_{\textrm{eff}}|n\rangle\langle n|\mathcal{H}_{\textrm{eff}}|M_\mu\rangle}{m_M-E_{n}+\mathrm{i}\epsilon}\notag \\
    &\equiv M'-\mathrm{i}\Gamma'/2\,.
  \end{align}
  where $\mathcal{H}_{\textrm{eff}}$ represents the effective Hamiltonian and $|n\rangle$ represents possible intermediate states. The oscillation probability from $M_\mu$ to $\overline{M_\mu}$ can be computed to be~\cite{Fukuyama:2021iyw}
  \begin{align}
    P(M_\mu^{i}\to \overline{M_\mu}^{i}) & =2|\mathcal{M}_{12}^{i}|^2/\Gamma_{M}^2\xlongequal{\Gamma'\rightarrow 0}\left|{\langle\overline{M}_\mu^i|\mathcal{H}_\textrm{eff}|M_\mu^i\rangle}\right|^2/ \left(2 m_M^2\Gamma_{M}^2\right)\,,
  \end{align}
  where $\Gamma'\rightarrow 0$ gives the leading order term and $i=P,V$ represents the para- and ortho-muonium states respectively. To apply experimental constraints, the spin-averaged oscillation probability is~\cite{Conlin:2020veq}
  \begin{equation}
    P^{\textrm{th}}_{M_\mu-\overline{M}_\mu}=\sum_{i=P,V}\frac{1}{2S_{i}+1}P(M_\mu^{i}\to \overline{M_\mu}^{i})< P^{\textrm{lim}}_{M_\mu-\overline{M}_\mu}/S_{B}\,,
  \end{equation}
  where $S_{P}=0,S_{V}=1$, $P^{\textrm{lim}}_{M_\mu-\overline{M}_\mu}$ is the limit from experiment and $S_{B}=0.95$~\cite{Willmann:1998gd} is a magnetic suppression factor. In our toy model, the effective Hamiltonian contributing to muonium-antimuonium oscillation is
  \begin{equation}
    \mathcal{H}_{\textrm{eff}}=\frac{Y_{e\mu}^2}{M_{\phi}^2}\bar\mu e\bar\mu e~,
  \end{equation}
  which yields the matrix elements~\cite{Conlin:2020veq}:
  \begin{equation}
    \langle\overline{M_\mu}^{P}|\bar\mu e\bar\mu e|M_\mu^{P}\rangle=-\frac{1}2f_{M}^2m_M^2,\quad \langle\overline{M_\mu}^{V}|\bar\mu e\bar\mu e|M_\mu^{V}\rangle=-\frac{3}2f_{M}^2m_M^2\,,
  \end{equation}
  where
  \begin{equation}
    f_{M}^2=\frac{4\alpha^{3}}{\pi m_M}\left(\frac{m_{e}m_{\mu}}{m_{e}+m_{\mu}}
    \right)^{3}=\frac{4(\alpha m_{e}')^{3}}{\pi m_M}~,
  \end{equation}
  with $\alpha$ the fine structure constant and $m_{e}'$ the reduced mass of electron in the muonium system. Therefore, the $M_\mu-\overline{M_\mu}$ oscillation probability is
  \begin{equation}
    P^{\textrm{th}}_{M_\mu-\overline{M}_\mu}=\frac{m_M^2f_{M}^{4}Y_{e\mu}^4}{2\Gamma_{M}^2M_\phi^4}=\frac{8Y_{e\mu}^{4}(\alpha m_{e}')^{6}}{\pi^2\Gamma_{M}^2M_{\phi}^{4}}\,.
  \end{equation}

  \subsubsection{Three-lepton LFV Decays}
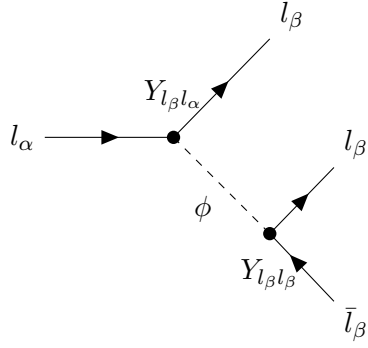
\begin{figure}[htbp]
    \centering
    \begin{tikzpicture}[baseline=(initial.base)]
      \begin{feynman}
        \vertex (initial) {$l_{\alpha}$}; 
        \vertex [dot, right=2.0cm of initial](Yab); 
        \vertex [above right=1.8cm of Yab] (lb) {$l_{\beta}$}; 
        \vertex [dot, below right=1.8cm of Yab] (Ycd); 
        \vertex [above right=1.2cm of Ycd] (lc){$l_{\beta}$}; 
        \vertex [below right=1.2cm of Ycd] (ld) {$\bar{l}_{\beta}$};

        \diagram*{ (initial) -- [fermion] (Yab), 
            (Yab) -- [fermion] (lb), 
            (Yab) -- [scalar, edge label'=$\phi$] (Ycd), 
            (Ycd) -- [fermion] (lc), (Ycd) -- [anti fermion] (ld)};

        \fill[black] (Yab) circle (2.5pt); 
        \fill[black] (Ycd) circle (2.5pt);
        \node[above=0.2cm of Yab] {$Y_{l_\beta l_\alpha}$}; 
        \node[below=0.2cm of Ycd] {$Y_{l_\beta l_\beta}$};
      \end{feynman}
    \end{tikzpicture}
    \caption{Diagram for 3-lepton LFV decay via scalar exchange.}
    \label{fig:l23l}
  \end{figure}

  As shown in Fig.~\ref{fig:l23l}, the three-lepton LFV decays $\ell_{\alpha} \to\ell_{\beta}\ell_{\beta}\bar{\ell}_{\beta}$ proceed at tree level through the exchange of $\phi$ and depend on the product of one off-diagonal Yukawa times one diagonal Yukawa $Y_{\ell_{\beta}\ell_{\alpha}}\,Y_{\ell_{\beta}\ell_{\beta}}$. The muon channel $\mu\to eee$ thus probes $Y_{e\mu}\,Y_{ee}$, complementary to the $Y_{e\mu}$-only information from $M_\mu-\overline{M}_\mu$ oscillation.

  From the toy model, the relevant effective 4-lepton interaction is given by
  \begin{equation}
    \mathcal{L}_{\textrm{eff;3-l}}=-\frac{Y_{\ell_\beta \ell_\alpha}Y_{\ell_\beta \ell_\beta}}{M_{\phi}^2}\bar{\ell}_{\beta}\ell_{\alpha}\bar{\ell}_{\beta}\ell_{\beta}\,.
  \end{equation}
  This gives the decay width of $\ell_{\alpha}\to \ell_{\beta}\ell_{\beta}\bar{\ell}
  _{\beta}$~\cite{Porod:2014xia}:
  \begin{equation}
    \Gamma_{\ell_\alpha \to \ell_\beta \ell_\beta \bar{\ell}_\beta}\approx \frac{m_{\alpha}^{5}}{2048 \pi^{3}M_{\phi}^{4}}Y_{\ell_\beta \ell_\alpha}^2Y_{\ell_\beta \ell_\beta}^2\,,
  \end{equation}
  and hence the branching ratio
  \begin{equation}
    \textrm{Br}_{\ell_\alpha \to \ell_\beta \ell_\beta \bar{\ell}_\beta}^{\textrm{th}}=\frac{\Gamma_{\ell_\alpha \to \ell_\beta \ell_\beta \bar{\ell}_\beta}}{\Gamma_{\ell_\alpha}}\,.
  \end{equation}
  Additionally, the branching ratios of the other two 3-lepton decay modes of tau lepton, $\tau\to e e\mu$ and $\tau\to \mu \mu e$, can be obtained numerically by Mathematica with packages FeynRules~\cite{Alloul:2013bka}, FeynArts~\cite{Hahn:2000kx}, and FormCalc~\cite{Hahn:2016ebn},
  \begin{align}
    \textrm{Br}^\textrm{th}_{\tau\to e e \mu}&=0.4294 \times Y_{e e}^2 Y_{\mu\tau}^2 + 0.3515 \times Y_{e\mu}^2 Y_{e\tau}^2 - 0.2147 \times Y_{e e} Y_{e\mu} Y_{e\tau} Y_{\mu\tau},\label{eq:eeu}\\
    \textrm{Br}^\textrm{th}_{\tau\to \mu\mu e}&=0.3326 \times Y_{\mu\mu}^2 Y_{e\tau}^2 + 0.4178 \times Y_{e\mu}^2 Y_{\mu\tau}^2 - 0.1663 \times Y_{\mu\mu} Y_{e\mu} Y_{e\tau} Y_{\mu\tau}\label{eq:uue}.
  \end{align}
  These two modes can provide supplementary constraints of combined effects from $Y_{e\mu}$, $Y_{e\tau}$ and $Y_{\mu\tau}$.
  
  \subsubsection{Radiative LFV Decays}

\begin{figure}[htbp]
    \centering
    \begin{tikzpicture}[baseline=(initial.base)]
      \begin{feynman}
        \vertex (initial) {$\ell_{\alpha}$}; 
        \vertex [right=1.5cm of initial] (Yab);
        \vertex [above right=1.2cm of Yab] (qed); 
        \vertex [above right=1cm of qed](gamma) {$\gamma$}; 
        \vertex [right=1.8cm of Yab] (Ybc); 
        \vertex [right=1.5cm of Ybc] (lc) {$\ell_{\beta}$};

        \diagram* { (initial) -- [fermion] (Yab), 
        (Yab) --[fermion,bend left=45,edge label=$\tau$] (qed) 
              --[fermion,bend left=45,edge label=$\tau$] (Ybc), 
        (Yab) --[scalar, edge label'=$\phi$] (Ybc) -- [fermion] (lc),
        (qed) --[photon] (gamma)};

        \fill[black] (Yab) circle (2.5pt); 
        \fill[black] (Ybc) circle (2.5pt); 
        \node[below=0.2cm of Yab] {$Y_{\ell_\alpha \tau}$}; 
        \node[below=0.2cm of Ybc] {$Y_{\ell_\beta \tau}$};
      \end{feynman}
    \end{tikzpicture}
    \caption{Leading order diagram for scalar mediated radiative LFV decay.}
    \label{fig:l2lg}
  \end{figure}
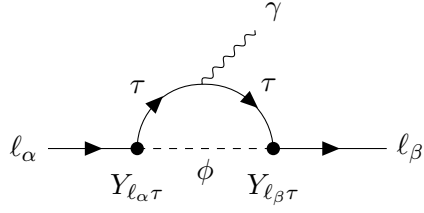

  As shown in Fig.~\ref{fig:l2lg}, the radiative LFV decays $\ell_{\alpha}\to\ell_{\beta}\gamma$ arise at one loop through a $\phi$ exchange with a charged lepton running in the loop. With hierarchical diagonal Yukawas, the $\tau$ loop provides the dominant contribution to $\mu\to e\gamma$, and the rate therefore probes the product $Y_{e\tau}\,Y_{\mu\tau}$; this is the coupling combination against which MEG~II sets the strongest bound in the global fit.

  Adopting an effective field theory approach, the leading order contribution to the decay width of $\ell_{\alpha}\to \ell_{\beta}\gamma$ is given by~\cite{Blankenburg:2012ex}
  \begin{equation}
    \Gamma^{\textrm{LO}}_{\ell_\alpha\to \ell_\beta\gamma}= \frac{ m_{\alpha}^{3}e_{c}^2}{8\pi}|A_{\alpha\beta}|^2\,,
  \end{equation}
  where the coefficients are
  \begin{align}
    A_{\mu e}&=\frac{1}{(4\pi)^2}Y_{e\tau}Y_{\mu\tau}\frac{m_{\tau}}{M_{\phi}^2}\left(\log{\frac{M_{\phi}^2}{m_{\tau}^2}}-\frac{3}2\right)\,,   \\
    A_{\tau l}&=\frac{1}{(4\pi)^2}Y_{\tau\tau}Y_{\ell\tau}\frac{m_{\tau}}{M_{\phi}^2}\left(\log{\frac{M_{\phi}^2}{m_{\tau}^2}}-\frac{4}{3}\right)\,.
  \end{align}
  Based on this, the branching ratio is approximately given by
  \begin{equation}
    \textrm{Br}_{\ell_\alpha\to \ell_\beta\gamma}^{\textrm{th}}  \approx \frac{\Gamma^{\textrm{LO}}_{\ell_\alpha\to \ell_\beta\gamma}}{\Gamma_{\ell_{\alpha}}}\,.
  \end{equation}

  \subsubsection{$\tau$ Semi-leptonic LFV Decays}
  Semileptonic $\tau$ decays $\tau\to\ell\,\pi^{+}\pi^{-}$ arise through the same $\bar{\ell}\tau\,GG$ operator that mediates $\mu-e$ conversion, now with the muon replaced by a tau and the hadronic matrix element evaluated at a momentum transfer in the di-pion system rather than at a nucleon. These decays therefore probe $\kappa_{gg}\,Y_{\ell\tau}$, providing a direct tau-sector counterpart to the muon-sector $\mu-e$ conversion constraint on $\kappa_{gg}\,Y_{e\mu}$.

  From the toy model the effective interaction can be written as
  \begin{equation}
    \mathcal{L}_{\textrm{eff;semi}}= -\frac{Y_{\ell\tau }\kappa_{gg}}{M_{\phi}^2}\bar{\ell}\tau G_{a}^{\mu\nu}G^{a}_{\mu\nu}+\textrm{h.c.}\,,
  \end{equation}
  from which one can derive the branching ratio~\cite{Cai:2018cog}:
  \begin{equation}
    \textrm{Br}_{\tau \to \ell \pi^+ \pi^-}^{\textrm{th}}=\frac{1}{\Gamma_{\tau}} \int_{4m_\pi^2}^{m_\tau^2}dq^2\,\frac{m_{\tau}}{648 \sqrt2\pi G_{F}}\left|{\frac{2Y_{\ell\tau }\kappa_{gg}}{\alpha_{s}v M_{\phi}^2}}\right|^2q^{4}\sqrt{1-\frac{4m_{\pi}^2}{q^2}}\left(1-\frac{q^2}{m^2_\tau}\right)^2\,
  \end{equation}
  where $\alpha_s=g_s^2/4\pi$ is the strong coupling constant and $v=246$ GeV is the electroweak vacuum expectation value.
  
  \subsubsection{Experimental Bounds for Low Energy cLFV Processes}

  We summarize the current experimental bounds on the cLFV modes discussed in this section in Table~\ref{tab:LFV_limits}. The projected sensitivity of corresponding future experiments are also presented.
  \begin{table}[htbp]
  \centering
  \renewcommand{\arraystretch}{1.2}
  \begin{tabular}{ccc}
    \toprule 
    \textbf{cLFV Process} & \textbf{Current Limit ($90\%$ C.L.)} & \textbf{Expected Future Sensitivity} \\
    \midrule 
    \multirow{2}{*}{\begin{tabular}{@{}l@{}}$\mu-e$ conversion \\ (capture ratio)\end{tabular}} 
      & \multirow{2}{*}{$7\times 10^{-13}$ (SINDRUM~II, Au~\cite{SINDRUMII:2006dvw})} 
      & $6.2\times 10^{-16}$ (Mu2e, Al~\cite{Miscetti:2025uxk}) \\
      & & $\sim 10^{-17}$ (COMET, Al~\cite{Moritsu:2022lem}) \\
    \midrule
    $\mu \to e\gamma$ 
      & $3.1\times 10^{-13}$ (MEG~II~\cite{MEGII:2023ltw}) 
      & $6\times 10^{-14}$ (MEG~II~\cite{Chiappini:2023luv}) \\
    \midrule
    $\mu \to e e \bar{e}$ 
      & $1.0 \times 10^{-12}$ (SINDRUM~\cite{SINDRUM:1987nra}) 
      & $\sim 10^{-16}$ (Mu3e~\cite{Hesketh:2022wgw}) \\
    \midrule
    \begin{tabular}{@{}c@{}}$M_\mu - \overline{M_\mu}$ oscillation \\ (probability)\end{tabular} 
      & $8.2\times 10^{-11}$ (MACS~\cite{Willmann:1998gd}) 
      & $\sim 10^{-13}$ (MACE~\cite{Bai:2024skk}) \\
    \midrule 
    \multirow{2}{*}{$\tau \to e e \bar{e}$} 
      & \multirow{2}{*}{$2.5\times 10^{-8}$ (Belle~II~\cite{Belle-II:2025urb})} 
      & $4.7\times 10^{-10}$ (Belle~II~\cite{Banerjee:2022vdd}) \\
      & & $1.9\times 10^{-10}$ (STCF~\cite{Achasov:2023gey}) \\
    \midrule
    $\tau \to e \mu \bar{e}$ 
      & $1.6\times 10^{-8}$ (Belle~II~\cite{Belle-II:2025urb}) 
      & $2.9\times 10^{-10}$ (Belle~II~\cite{Banerjee:2022vdd}) \\
    \midrule
    $\tau \to e e \bar{\mu}$ 
      & $1.6\times 10^{-8}$ (Belle~II~\cite{Belle-II:2025urb}) 
      & $2.3\times 10^{-10}$ (Belle~II~\cite{Banerjee:2022vdd}) \\
    \midrule
    \multirow{2}{*}{$\tau \to e\gamma$} 
      & \multirow{2}{*}{$5.6\times 10^{-8}$ (Belle~\cite{Belle:2021ysv})} 
      & $9.0\times 10^{-9}$ (Belle~II~\cite{Banerjee:2022vdd}) \\
      & & $5.7\times 10^{-9}$ (STCF~\cite{Achasov:2023gey}) \\
    \midrule
    \multirow{2}{*}{$\tau \to e \pi^{+}\pi^{-}$} 
      & \multirow{2}{*}{$2.3\times 10^{-8}$ (Belle~\cite{Belle:2021ysv})} 
      & $5.8\times 10^{-10}$ (Belle~II~\cite{Banerjee:2022vdd}) \\
      & & $\sim 10^{-10}$ (STCF~\cite{Achasov:2023gey}) \\
    \midrule 
    \multirow{3}{*}{$\tau \to \mu\mu\bar{\mu}$} 
      & $1.9\times 10^{-8}$ (Belle~II~\cite{Belle-II:2024sce}) 
      & $3.6\times 10^{-10}$ (Belle~II~\cite{Banerjee:2022vdd}) \\
      & $1.9\times 10^{-8}$ (LHCb~\cite{LHCb:2026eod}) 
      & $1.9\times 10^{-10}$ (STCF~\cite{Achasov:2023gey}) \\
      & $8.7\times 10^{-8}$ (ATLAS~\cite{ATLAS:2026kkw}) 
      & --- \\
    \midrule 
    $\tau \to \mu e \bar{\mu}$ 
      & $2.4\times 10^{-8}$ (Belle~II~\cite{Belle-II:2025urb}) 
      & $4.5\times 10^{-10}$ (Belle~II~\cite{Banerjee:2022vdd}) \\
    \midrule
    $\tau \to \mu \mu \bar{e}$ 
      & $1.3\times 10^{-8}$ (Belle~II~\cite{Belle-II:2025urb}) 
      & $2.6\times 10^{-10}$ (Belle~II~\cite{Banerjee:2022vdd}) \\
    \midrule
    \multirow{2}{*}{$\tau \to \mu\gamma$} 
      & \multirow{2}{*}{$4.2\times 10^{-8}$ (Belle~\cite{Belle:2021ysv})} 
      & $6.9\times 10^{-9}$ (Belle~II~\cite{Banerjee:2022vdd}) \\
      & & $5.7\times 10^{-9}$ (STCF~\cite{Achasov:2023gey}) \\
    \midrule
    \multirow{2}{*}{$\tau \to \mu \pi^{+}\pi^{-}$} 
      & \multirow{2}{*}{$2.1\times 10^{-8}$ (Belle~\cite{Belle:2021ysv})} 
      & $5.6\times 10^{-10}$ (Belle~II~\cite{Banerjee:2022vdd}) \\
      & & $\sim 10^{-10}$ (STCF~\cite{Achasov:2023gey}) \\
    \bottomrule
  \end{tabular}
  \caption{Current limits and expected future sensitivities for the cLFV processes considered in this work.}
  \label{tab:LFV_limits}
\end{table}

We also verify in App.~\ref{app:Z} that the constraints from one-loop contributions to $Z$ LFV decays are negligible compared to the constraints derived in this section.

  \section{Analysis and Results}
  \label{sec:analysis}

  \subsection{Set up of MCMC Analysis}
  \label{sec:mcmc-setup}

  The major result of this work is delivered through a Bayesian approach, focusing on the viable parameter space given the various experimental and theoretical constraints presented in Sec.~\ref{sec:observables}. Therefore, a Markov Chain Monte Carlo (MCMC) analysis is performed to explore the parameter space of this toy model, with the free parameters $\kappa_{gg}$,$Y_{e\mu}$,$Y_{e\tau}$, $Y_{\mu\tau}$,$Y_{ee}$,$Y_{\mu\mu}$, and $Y_{\tau\tau}$. We use package Zeus~\cite{karamanis2021zeus,karamanis2020ensemble} to perform the MCMC process and the visualization of the posterior distribution is managed through the corner package~\cite{corner}.

  Almost all physical observables discussed in Sec.~\ref{sec:observables} involve only the squares of the Yukawa couplings, so the sign of each $Y_{ij}$ is largely irrelevant.\footnote{The only exceptions are the two mixing terms in Eq.~\eqref{eq:uue} and~\eqref{eq:eeu}. However, as the subdominant mixing terms, they do not change the overall likelihood in a noticeable way.} For simplicity we denote $Y_{ij}$ as the corresponding absolute values. We take their prior distributions to be log-uniform distributions, i.e. the MCMC sampling variables are defined as 
  \begin{equation}
  x=(\log_{10}(\kappa_{gg}\textrm{GeV}),\log_{10}Y_{e\mu},\log_{10}Y_{e\tau},\log_{10}Y_{\mu\tau},\log_{10}Y_{ee},\\\log_{10}Y_{\mu\mu},\log_{10}Y_{\tau\tau})\,,
  \end{equation}
  reflecting maximal ignorance about the order of magnitude of the couplings and ensuring sensitivity across multiple decades. In the absence of theoretical guidance on the expected scale of these couplings, this choice is standard for positive parameters that can span several orders of magnitude. The prior distributions are assumed to be uncorrelated for simplicity. The upper and lower limits of each dimension of $x$ are taken to be $-10$ and $0$, respectively.

  Aside from displaying the marginalized posterior distributions, the credible intervals of each $x$ component are also reported. For the BSM couplings other than $\kappa_{gg}$ and $Y_{e\mu}$, which are not directly constrained by the CMS excess, the marginalized posterior distributions typically extend down to the lower prior boundary without vanishing. Their corresponding 95\% credible intervals are therefore one-sided, and we quote only upper limits. Because such one-sided limits depend on the choice of the prior's lower bound, a mild prior dependence of these values is introduced. A discussion of this prior sensitivity is presented in App.~\ref{app:prior}, in which we also comment on the analogous dependence of two-dimensional credible regions.

  The corresponding likelihood of a particular choice of parameter $x$ is described by the uncorrelated sum of the corresponding $\chi^2$ values from each experiment, namely:
  \begin{equation}
    \chi^2(x)=\chi^2_{146}(x)+\chi^2_{\mu\mu}(x)+\chi^2_{\tau\tau}(x)+\chi^2_{\textrm{LFV}}(x)~.
  \end{equation}
  The $\chi_{146}^2$ term comes from the direct LHC observation of the $146~\mathrm{GeV}$ signal in the $e\mu$ channel~\cite{CMS:2023pte}. We adopt the $2.8\sigma$ global significance and its corresponding $3.89\textrm{fb}$ cross section for the analysis, which reads:
  \begin{equation}
    \sigma_{pp\to \phi \to e\mu}^{\textrm{exp}}=3.89 \textrm{ fb}\,.\label{eq:signal}
  \end{equation}
  Therefore, the corresponding $\chi^2$ term takes the form:
  \begin{equation}
    \label{eq:chi146}\chi^2_{\textrm{146}}(x)=\left[\frac{\sigma_{pp\to \phi \to e\mu}^{\textrm{th}}(x)-\sigma_{pp\to \phi \to e\mu}^{\textrm{exp}}}{\sigma_{pp\to \phi \to e\mu}^{\textrm{exp}}/2.8}\right]^2\,,
  \end{equation}
  On the other hand, the $\chi^2_{\mu\mu}(x)$ and $\chi^2_{\tau\tau}(x)$ terms come from the same flavor di-lepton resonance searches~\cite{CMS:2019mij,CMS:2022goy} at the LHC, with no BSM resonance found. Due to the large SM backgrounds, their LHC upper limits are significantly larger, reaching $\mathcal{O}$(pb). Yet searches for $e e$ final state at the LHC~\cite{ATLAS:2019erb,CMS:2019kaf,CMS:2026cjr} did not cover the $146~\textrm{GeV}$ benchmark here, so the corresponding constraints are not included in this work. The 95\% C.L. upper limits for $\sigma^{\textrm{lim.}}_{pp\to \phi \to \textrm{X}}, ~\textrm{X}=\mu\mu(\tau\tau)$ at $M_\phi=146\textrm{ GeV}$ through ggF are 225(800)~fb, respectively. The corresponding $\chi^2$ terms are thus:
  \begin{equation}
    \chi^2_{X}(x)= \left[\frac{\sigma_{pp\to \phi \to X}^{\textrm{th}}(x)-0}{\sigma_{pp\to \phi \to X}^{\textrm{lim}}/1.96}\right]^2\,,
  \end{equation}
  where the factor of $1.96$ stands for the standard deviation needed for a one-sided Gaussian distribution for the $95\%$ C.L. upper limits. 

  The $\chi^2$ terms from the cLFV processes in Table~\ref{tab:LFV_limits} are constructed analogously, assuming that they are Gaussian and independent:
  \begin{align}
    \chi^2_{\textrm{LFV}}(x)=\sum_{\substack{\textrm{process }i\\\textrm{in Table \ref{tab:LFV_limits}}}}\chi^2_{i}(x)=\sum_{\substack{\textrm{process }i\\\textrm{in Table \ref{tab:LFV_limits}}}}\left[\frac{\textrm{Br}^{\textrm{th}}_{i}(x)-0}{\textrm{Br}^{\textrm{lim}}_{i}/1.645}\right]^2\,,
  \end{align}
  where $\textrm{Br}^\textrm{th}_i$s are given in Sec.~\ref{sec:LFV} and the factor of $1.645$ stands for the standard deviation needed for a one-sided Gaussian distribution for the $90\%$ C.L. upper limits. For the branching ratios of $\tau\to ee\mu$ in Eq.~\eqref{eq:eeu} and $\tau\to\mu\mu e$ in Eq.~\eqref{eq:uue}, we drop the interference terms and retain only the incoherent sum of the two squared amplitudes for simplicity. For comparison, some of the projected experimental reach in the near future are also presented in Table~\ref{tab:LFV_limits}. For the $\mu$ sector, the constraints are dominated by precision muon experiments, while the $\tau$ sector constraints are dominated by collider experiments.

  \subsection{Numerical Results}
  \begin{figure}[htbp]
    \includegraphics[width=\textwidth]{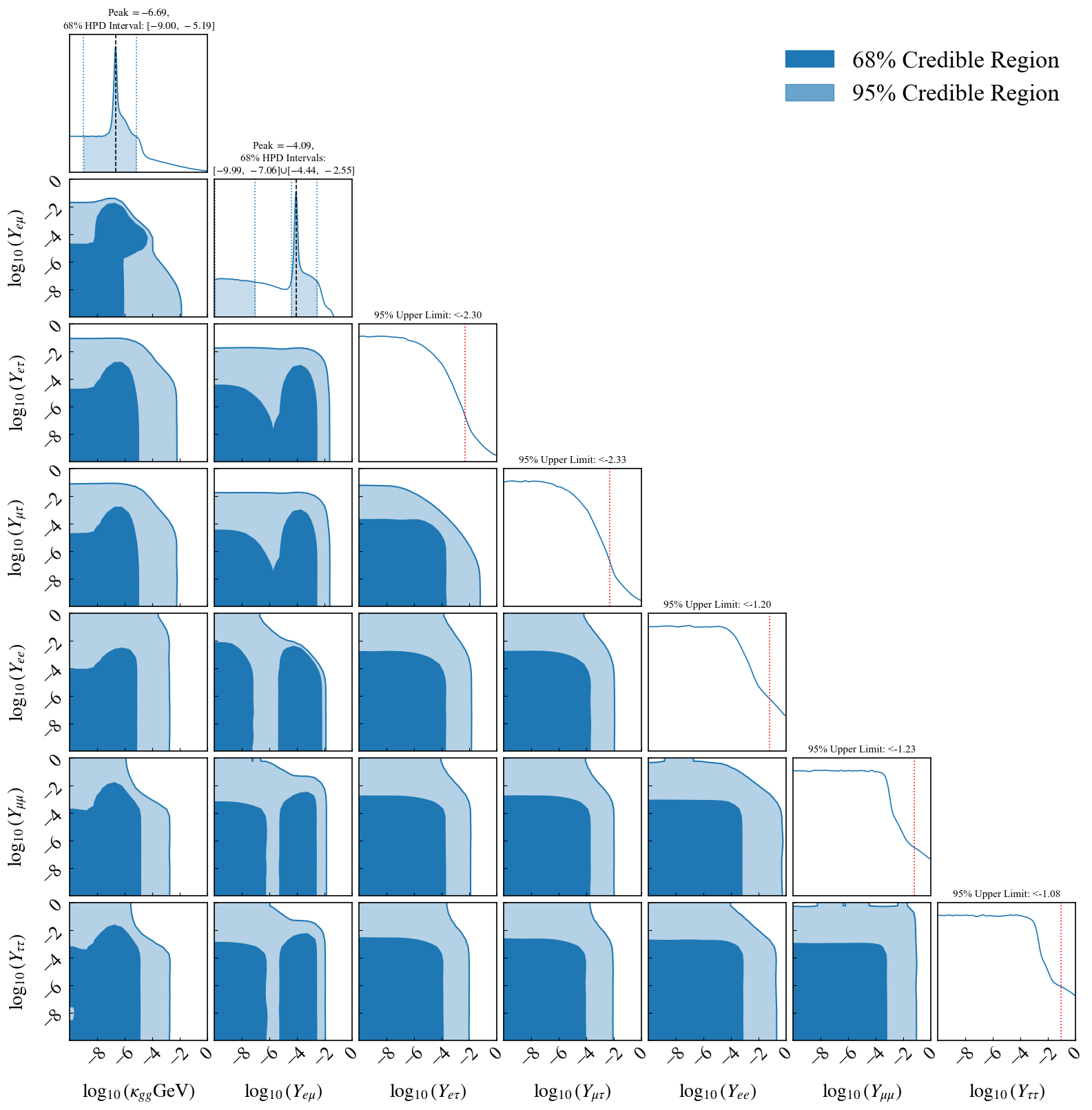}
    \centering
    \caption{Numerical results of the MCMC analysis as a corner plot, with 7 parameters in total. In the non-diagonal panels, the 2D marginalized posterior distributions are shown, with the darker(lighter) shades standing for the $68(95)\%$ credible regions. In the diagonal panels, the marginalized 1D distributions are shown.
    The corresponding likelihood peak values and 68\% HPD intervals (shaded 1D region) for $\log_{10}(\kappa_{gg}\textrm{GeV})$ and $\log_{10}(Y_{e\mu})$ are reported on their 1D panel, respectively. A 95\% credible upper bound is reported for each other parameters on the 1D panel.}
    \label{fig:corner}
  \end{figure}

  \begin{table}[htbp]
  \centering
  \renewcommand{\arraystretch}{1.3}
   \begin{tabular}{ccc}
    \toprule
    \textbf{Parameter} & \textbf{Constraint Type} & \textbf{Posterior Value} \\
    \midrule
   $\log_{10}(\kappa_{gg}\textrm{GeV})$       & 68\% HPD Interval   & $[-9.00, -5.19]$ \\
   $\log_{10}(Y_{e\mu})$     & 68\% HPD Intervals   & $[-9.99,-7.06]\cup [-4.44, -2.55]$ \\
   \midrule
   $\log_{10}(Y_{e\tau})$       & 95\% Upper Limit    & $< -2.31$ \\
   $\log_{10}(Y_{\mu\tau})$   & 95\% Upper Limit    & $< -2.33$ \\
   $\log_{10}(Y_{ee})$ & 95\% Upper Limit    & $< -1.20$ \\
   $\log_{10}(Y_{\mu\mu})$    & 95\% Upper Limit    & $< -1.23$ \\
   $\log_{10}(Y_{\tau\tau})$  & 95\% Upper Limit    & $< -1.08$ \\
   \bottomrule
   \end{tabular}
   \caption{Summary of posterior constraints on the model parameters in $\log_{10}$ scale.}
   \label{tab:mcmc_constraints}
   \end{table}
   
  The marginalized posterior distribution resulting from the MCMC analysis is presented in Fig.~\ref{fig:corner} as a corner plot. The two-dimensional panels below the diagonal show the marginalized posterior co-distribution for each pair of parameters, with the darker and lighter shaded regions corresponding to the $1\sigma$ ($68\%$) and $2\sigma$ ($95\%$) credible regions, respectively. The diagonal panels show the marginalized one-dimensional posteriors for each parameter. For the couplings that receive direct constraints from the CMS excess, i.e. $\kappa_{gg}$ and $Y_{e\mu}$, the maximum-likelihood values and the $68\%$ credible highest probability density (HPD) intervals~\cite{Chen01031999} are quoted in the appropriate places. The two disjoint credible regions for $Y_{e\mu}$ reflect two competing tendencies: the first, which extends to the lower prior boundary, is driven by the null results of low-energy cLFV searches; the second, featuring a distinct peak around $\log_{10}(Y_{e\mu}) \sim -4.09$, accommodates the CMS excess. In contrast, $\kappa_{gg}$ exhibits a unimodal 68\% credible interval with its posterior maximum at $\log_{10}(\kappa_{gg},\mathrm{GeV}) \sim -6.69$.". The details about the HPD intervals are discussed in App.~\ref{app:HPD}. For the remaining couplings, only the $95\%$ upper limits of the credible interval are given, as discussed in Sec.~\ref{sec:mcmc-setup}. All of those posterior constraint values are listed in Table~\ref{tab:mcmc_constraints}. From the analysis, the upper limits for LFV couplings other than $Y_{e\mu}$ are about $10^{-2}$, while the corresponding values for diagonal couplings are weaker by an order of magnitude. Although the di-lepton resonance searches, $pp\to\phi\to\mu\mu$~\cite{CMS:2019mij} and $pp\to\phi\to\tau\tau$~\cite{CMS:2022goy},  enter the likelihood through $\chi^2_{\mu\mu}$ and $\chi^2_{\tau\tau}$, they do not provide strong constraints here. As reported in Table~\ref{tab:mcmc_constraints}, all three diagonal Yukawas $Y_{ee}$, $Y_{\mu\mu}$, and $Y_{\tau\tau}$ share comparable 95\% upper limits, though only $Y_{\mu\mu}$ and $Y_{\tau\tau}$ are directly constrained by the resonance searches.
  
  From Fig.~\ref{fig:corner}, the two-dimensional distribution of $\kappa_{gg}$ and $Y_{e\mu}$ shows a clear bimodal structure. Although the $1\sigma$ and $2\sigma$ contours appear connected, the posterior probability density peaks at two distinct regions with significant support. The first mode is driven by the evidence for a $146\,\mathrm{GeV}$ resonance, in which both $\kappa_{gg}$ and $Y_{e\mu}$ are relatively large, and a negative correlation is visible because their product is directly related to the observed excess rate. The second mode populates extremely small values of $\kappa_{gg}$ and $Y_{e\mu}$ (the tiny-coupling regime), favored by the non-observation of additional di-lepton resonances and by the low energy cLFV constraints, even though it yields a larger $\chi^2_{146}$ term in Eq.~\eqref{eq:chi146}. The improvement in the other likelihood components compensates partially. In other words, the $2.8\sigma$ global significance is insufficient to outweigh all the low energy cLFV null observations and favor a sizable $Y_{e\mu}$. We expect future LHC analysis, either positive or negative, will change the current status.
  
  In the first mode, the dominant mechanism bounding the diagonal couplings is the total-width effect. The observed CMS excess requires $\kappa_{gg}^2 Y_{e\mu}^2 / \Gamma_\phi$ to be sizable, while the non-observation of $\mu-e$ conversion imposes a stringent upper bound on the product $\kappa_{gg} Y_{e\mu}$. Satisfying both constraints simultaneously forces $\Gamma_\phi$, which receives contributions from the squared diagonal and off-diagonal Yukawas, to remain small, thereby bounding all Yukawa couplings from above. In the second mode, however, both $\kappa_{gg}$ and the off-diagonal Yukawas are driven toward the lower prior boundary, so neither the CMS signal constraint nor the low-energy cLFV bounds effectively restrict the diagonal Yukawas, which then spread over the full prior range. Because the second mode carries substantial posterior weight, with the second mode carrying roughly 80\% of the posterior weight at the 68\% credible level, the marginalized upper bounds on the diagonal Yukawas remain rather loose.

\begin{figure}[htbp]
  \centering
  \includegraphics[width=\textwidth]{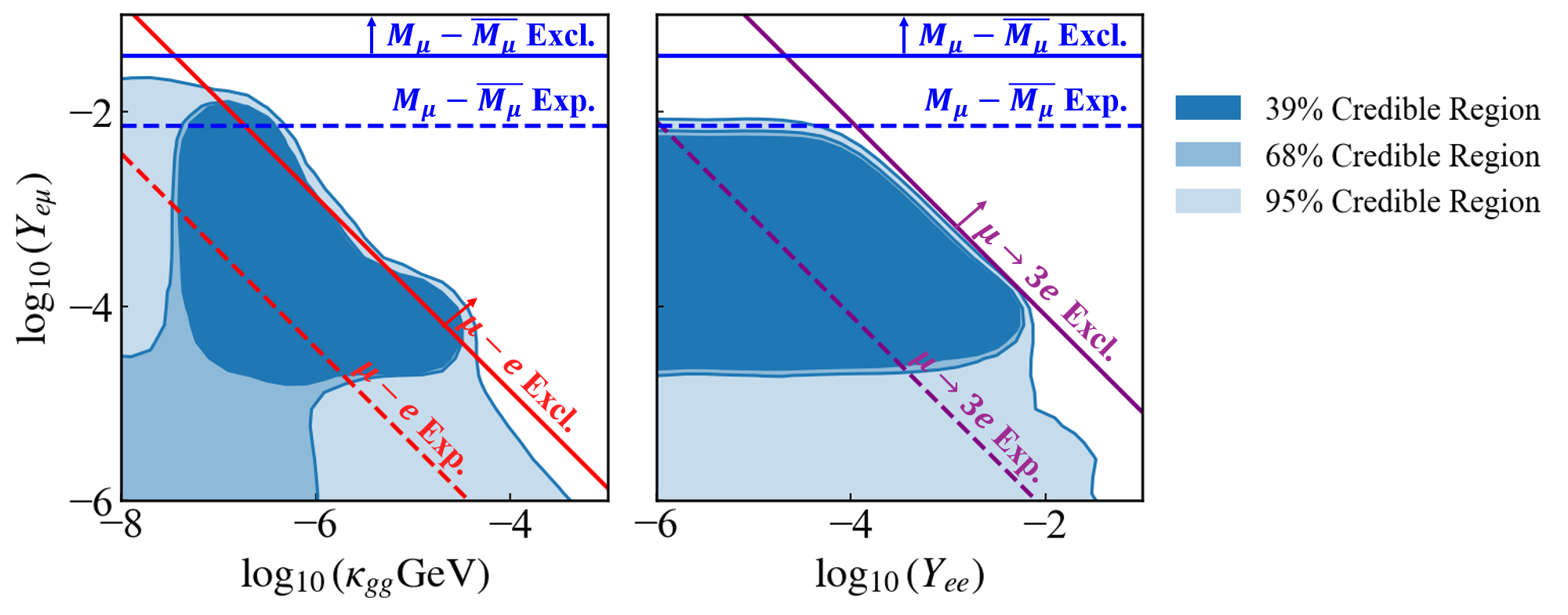}
  \caption{ Constraints for $Y_{e\mu}$. The two panels show the two-dimensional posterior in the ($\log_{10}\kappa_{gg}\textrm{GeV}$,$\log_{10}Y_{e\mu}$) (left) and ($\log_{10}Y_{ee}$,$\log_{10}Y_{e\mu}$) (right) planes. Both panels are overlaid with the current limits and projected sensitivities from $M_\mu\to\overline{M_\mu}$ oscillation; the left panel additionally includes those from $\mu-e$ conversion, and the right panel additionally includes those from $\mu\to eee$.}
  \label{fig:res_eu}
\end{figure}

\begin{figure}[htbp]
\centering
\includegraphics[width=\textwidth]{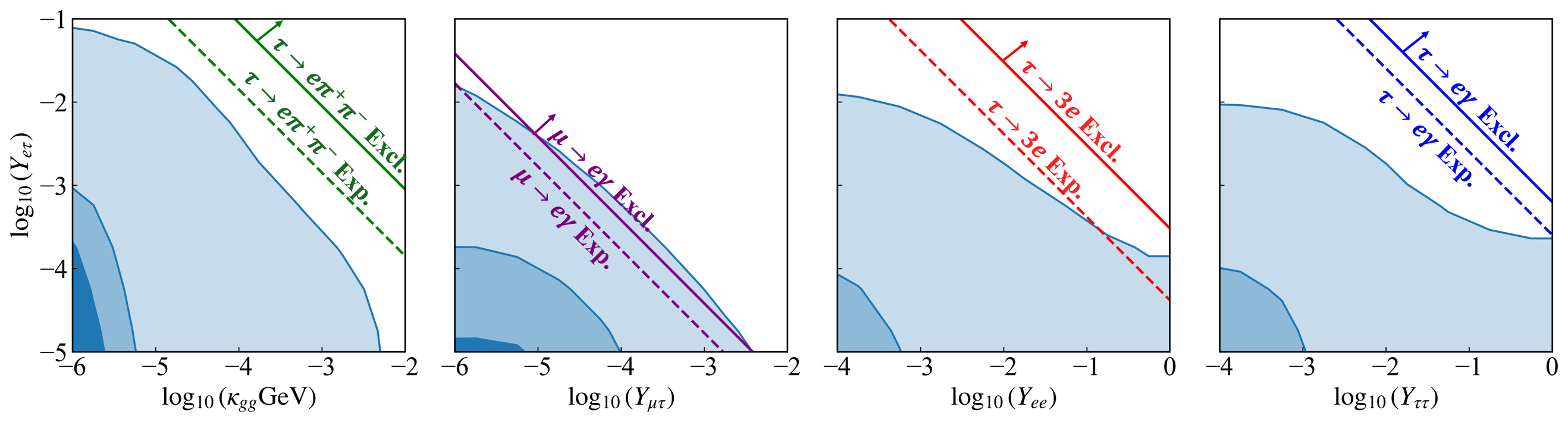}
\caption{Constraints for $Y_{e\tau}$. Shading of the credible regions follows the same pattern as in Fig.~\ref{fig:res_eu}. The four panels show the two-dimensional posterior in the $\log_{10}Y_{e\tau}$ plane against $\log_{10}(\kappa_{gg}\textrm{GeV})$, $\log_{10}Y_{\mu\tau}$, $\log_{10}Y_{ee}$, and $\log_{10}Y_{\tau\tau}$, overlaid with the current limits and expected sensitivities from $\tau\to e\pi^{+}\pi^{-}$, $\mu\to e\gamma$, $\tau\to eee$, and $\tau\to e\gamma$, respectively.}
\label{fig:res_et}
\end{figure}

\begin{figure}[htbp]
\centering
\includegraphics[width=\textwidth]{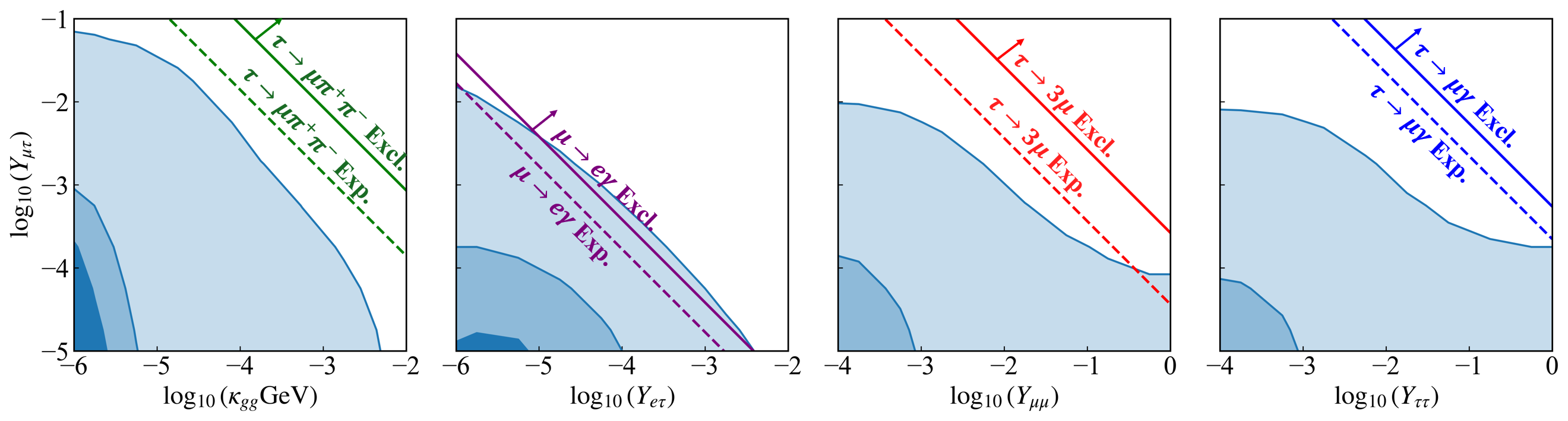}
\caption{Constraints for $Y_{\mu\tau}$. Shading of the credible regions follows the same pattern as in Fig.~\ref{fig:res_eu}. The four panels show the two-dimensional posterior in the $\log_{10}Y_{\mu\tau}$ plane against $\log_{10}(\kappa_{gg}\textrm{GeV})$, $\log_{10}Y_{e\tau}$, $\log_{10}Y_{\mu\mu}$, and $\log_{10}Y_{\tau\tau}$, overlaid with the current limits and expected sensitivities from $\tau\to\mu\pi^{+}\pi^{-}$, $\mu\to e\gamma$, $\tau\to\mu\mu\mu$, and $\tau\to\mu\gamma$, respectively.}
\label{fig:res_ut}
\end{figure}

  To better illustrate the physical origin of posterior distribution patterns and the prospects for future measurements, we examine selected two-dimensional planes in Fig.~\ref{fig:res_eu}, Fig.~\ref{fig:res_et} and Fig.~\ref{fig:res_ut}, overlaid with the current  limits and projected experimental sensitivities for the relevant cLFV processes listed in Table~\ref{tab:LFV_limits}. For example, Fig.~\ref{fig:res_eu} shows the $(\log_{10}(\kappa_{gg}\textrm{GeV}), \log_{10}Y_{e\mu})$ and $(\log_{10}Y_{ee}, \log_{10}Y_{e\mu})$ planes for the $e-\mu$ sector, together with contours of $\mu-e$ conversion, $M_\mu \to \overline{M_\mu}$ oscillation, and $\mu\to eee$ decay. In these panels, the branching-ratio predictions are drawn directly: solid lines correspond to the current $90\%$ C.L. experimental limits, and dashed lines to projected future sensitivities.

  As shown in Fig.~\ref{fig:res_eu}, the future sensitivity for the first mode will be driven primarily by $\mu-e$ conversion~\cite{Moritsu:2022lem,Miscetti:2025uxk} combined with $\mu\to eee$ decay~\cite{Hesketh:2022wgw}, rather than by $M_\mu- \overline{M_\mu}$ oscillation~\cite{Bai:2024skk}. In particular, the next-generation $\mu-e$ conversion experiments are expected to probe most of the parameter space favored by the $146\,\mathrm{GeV}$ excess, although a small region may still survive.  Similarly, Fig.~\ref{fig:res_et} and Fig.~\ref{fig:res_ut} show the analogous planes for the $e-\tau$ and $\mu-\tau$ sectors. In general, the global constraints from the Bayesian analysis are more restrictive than the individual cLFV bounds, incorporating all constraints simultaneously and marginalizing over the other model parameters. The only exception is the $(\log_{10} Y_{e\tau},\log_{10} Y_{\mu\tau})$ plane, where the indirect bound from $\mu\to e\gamma$ (mediated by a one-loop diagram with a $\tau$ lepton) is particularly stringent, shaping the credible region.
  
  Taken together, the MCMC posterior and the per-process exclusions confirm the complementarity anticipated in Sec.~\ref{sec:intro}. However, the evidence for the new resonance is not strong enough to exclude the possibility of vanishing cLFV couplings, especially once the look-elsewhere effect (global significance) is taken into account. The CMS observation favors a non-zero combination $\kappa_{gg}^2Y_{e\mu}^2/\Gamma_{\phi}$, while the low-energy cLFV experiments and the LHC same-flavor di-lepton resonance searches carve out the remaining directions and leave a parameter region without sizable cLFV. Future experiments will probe this region directly: Mu2e, COMET and Mu3e will measure $Y_{e\mu}$ within the next decade; $Y_{e\tau}$ and $Y_{\mu\tau}$ are constrained primarily through their product by $\mu\to e\gamma$, with MEG~II extending the reach; and $Y_{\mu\mu}$, $Y_{\tau\tau}$ will be further tightened by the HL-LHC~\cite{ATLAS:2019mfr} same-flavor channels.

  \section{Conclusion}
  \label{sec:conclusion} 
  The recent CMS hint of a $\sim 146~\mathrm{GeV}$ resonance decaying into $e\mu$, if confirmed, would constitute a major piece of evidence of charged lepton flavor violation. In this work, we examined which regions of scalar-mediator coupling space are simultaneously compatible with the claimed LHC signal and existing cLFV constraints in a model-agnostic manner. Answering this question maps out the present status and future sensitivity of the global experimental program for cLFV.

  To this end, we introduced a minimal effective description: a single real scalar $\phi$ of mass $146~\mathrm{GeV}$ coupled to gluons through a dimension-five operator and to all charged-lepton bilinears through symmetric Yukawa couplings, giving seven free parameters in total. A Bayesian MCMC analysis with log-uniform priors spanning ten orders of magnitude was used to explore the posterior distribution of these couplings against the current experimental constraints. The resulting posterior reveals a bimodal structure: the CMS excess favors a non-zero $Y_{e\mu}$ and therefore a genuine cLFV signal, yet a second mode with tiny cLFV couplings remains viable because the global significance of the excess is only $\sim 2.8\sigma$. Within the LHC-signal-favoring branch, $Y_{e\mu}$ is driven to a preferred non-zero value, and can be partially probed by the current $\mu-e$ conversion experiment. The $\tau$-sector off-diagonal Yukawas $Y_{e\tau}$ and $Y_{\mu\tau}$ are constrained primarily through their product by the one-loop contribution to $\mu\to e\gamma$, which currently sets a tighter bound than any direct $\tau$ LFV search. The LHC same-flavor di-lepton resonance searches provide supplemental constraints for 
  the diagonal couplings $Y_{\mu\mu}$ and $Y_{\tau\tau}$.
  
  The complementarity between high-energy and low-energy cLFV probes is quantitative: the LHC measurement fixes one coupling combination, the low-energy experiments carve out the remaining directions in the seven-parameter space, and the projected sensitivities align with the preferred region rather than merely improving existing exclusions. This structure is a generic feature of scalar-mediator interpretations of the excess and is robust against the choice of UV completion. We have worked throughout in the ggF production regime; a parallel analysis in the complementary lepton-PDF production scenario of Ref.~\cite{Afik:2023vit} is left for future work. Ultimately, the resolution of the $146~\mathrm{GeV}$ question will come from experiment, and the framework presented here identifies the coupling directions that each experimental program will constrain and provides a unified language for interpreting their combined reach.

  Future experiments will directly probe the parameter regions indicated by the posterior. The projected sensitivities of Mu2e, COMET, Mu3e, and MACE will cover the full signal-preferred range of $Y_{e\mu}$ within the next several years, thereby also resolving the two posterior modes. MEG~II will tighten the $\mu\to e\gamma$ constraint on the product $Y_{e\tau}Y_{\mu\tau}$ by an order of magnitude, while Belle~II and the proposed STCF will improve the direct $\tau$-sector probes. The HL-LHC with $3000~\mathrm{fb}^{-1}$ of integrated luminosity will extend the reach on the diagonal $Y_{\mu\mu}$ and $Y_{\tau\tau}$ couplings correspondingly. More broadly, the CMS $146~\mathrm{GeV}$ excess, if it reflects new physics, is a scenario that many independent experiments will address in parallel within the next decade. A heterogeneous dataset gathered from multiple independent experiments and frontiers in the future shall be able to confirm or exclude the scalar-mediator interpretation.

  \section*{Acknowledgment}
 We thank Innes Bigaran, Ryan Plestid's helpful discussions at the early stage of this work. C.G. and Z.J.X's work are supported by Natural Science Foundation of Guangdong Province (Grant No.2024QN11X110).

  \appendix

  \section{Evaluation of the gluonic form factor $f_{GN}$}
  \label{app:fGN} 
  We collect here the lattice-QCD and meson-physics inputs used to evaluate the gluonic form factor $f_{GN}$ of Eq.~\eqref{eq:fGN-def} that enters the $\mu-e$ conversion rate. The scalar form factors $f^{q}_{SN}$ for light quarks $q = u,d,s$ and nucleon $N = p,n$ are given by~\cite{Cirigliano:2009bz}
  \begin{align}
    f^{u}_{Sp} & = \frac{m_{u}}{m_{u}+m_{d}}\,(1+\xi)\,\frac{\sigma_{\pi N}}{m_{p}}, & f^{u}_{Sn} & = \frac{m_{u}}{m_{u}+m_{d}}\,(1-\xi)\,\frac{\sigma_{\pi N}}{m_{p}}, \\
    f^{d}_{Sp} & = \frac{m_{d}}{m_{u}+m_{d}}\,(1-\xi)\,\frac{\sigma_{\pi N}}{m_{p}}, & f^{d}_{Sn} & = \frac{m_{d}}{m_{u}+m_{d}}\,(1+\xi)\,\frac{\sigma_{\pi N}}{m_{p}}, \\
    f^{s}_{Sp} & = \frac{m_{s}}{m_{u}+m_{d}}\,y\,\frac{\sigma_{\pi N}}{m_{p}},       & f^{s}_{Sn} & = \frac{m_{s}}{m_{u}+m_{d}}\,y\,\frac{\sigma_{\pi N}}{m_{p}},
  \end{align}
  
  where
  
  \begin{equation}
    \sigma_{\pi N}= \frac{m_{u}+m_{d}}2\, \langle p|\bar{u}u + \bar{d}d|p\rangle
    ,\quad \xi = \frac{\langle p|\bar{u}u-\bar{d}d|p\rangle}{\langle p|\bar{u}u+\bar{d}d|p\rangle}
    ,\quad y = \frac{2\langle p|\bar{s}s|p\rangle}{\langle p|\bar{u}u+\bar{d}d|p\rangle}
    .
  \end{equation}
  
  We use the $N_{f}=2+1$ averages from FLAG 2024~\cite{FlavourLatticeAveragingGroupFLAG:2024oxs}:
  
  \begin{align}
    \sigma_{\pi N}                         & = 42.2 \pm 2.4~\mathrm{MeV},                                      \\
    \sigma_{s}                             & = m_{s}\,\langle p|\bar{s}s|p\rangle = 44.9 \pm 6.4~\mathrm{MeV}, \\
    \langle p|\bar{u}u - \bar{d}d|p\rangle & = 1.11^{+0.14}_{-0.16},                                          &
  \end{align}
  
  and quark mass terms from PDG 2024~\cite{ParticleDataGroup:2024cfk}:
  \begin{align}
    m_{u}/m_{d}   & = 0.462 \pm 0.020,                              \\
    \bar{m}       & = (m_{u}+m_{d})/2 = 3.49 \pm 0.07~\mathrm{MeV}, \\
    m_{s}/\bar{m} & = 27.33^{+0.18}_{-0.14}
  \end{align}
  Substituting into the scalar form factors and the definition $f_{GN}= 1 - \sum_{q=u,d,s}f^{q}_{SN}$ yields the numerical values quoted in Eq.~\eqref{eq:fGN-values},
  
  \begin{equation}
    f_{Gp}\approx 0.909,\qquad f_{Gn}\approx 0.906.
  \end{equation}
  
  The shift of $f_{GN}$ by taking the $N_{f}=2+1+1$ FLAG value is $<1\%$ and therefore has negligible effect on the conversion rate and on the bounds derived in Sec.~\ref{sec:analysis}.

\section{Prior Dependence}
  \label{app:prior}
  \begin{figure}[H]
    \includegraphics[width=\textwidth]{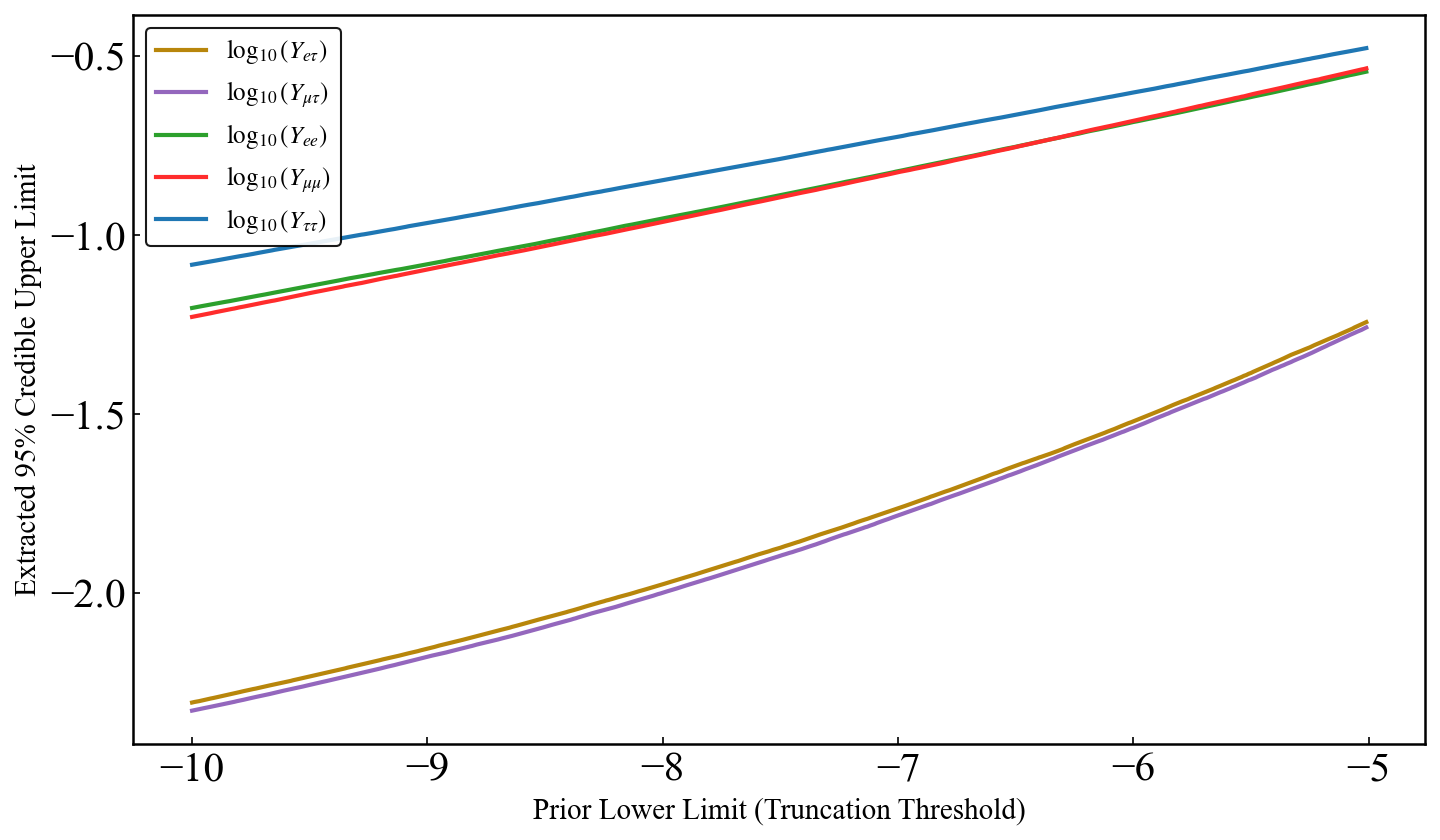}
    \centering
    \caption{The sensitivity of 95\% credible upper limit to the prior lower limit.}
    \label{fig:truncation}
  \end{figure}

  To assess the robustness of the upper limit against the choice of prior range, we perform a truncation sensitivity test upon the couplings of which only upper limits are given, i.e. $\log_{10}Y_{e\tau}$,$\log_{10}Y_{\mu\tau}$,$\log_{10}Y_{e e}$,$\log_{10}Y_{\mu\mu}$, and $\log_{10}Y_{\tau\tau}$. In this test, we vary the lower boundary of the prior on those couplings and re-extract the corresponding upper limit of the 95\% credible interval. The results are presented in Fig.~\ref{fig:truncation}, which shows that varying the lower bound by an order of magnitude shifts the upper limit by less than 30\%.

\section{Highest Posterior Density Intervals in Bayesian Analysis}
\label{app:HPD}

Roughly speaking, the HPD interval $[\theta_L,\,\theta_U]$ at credibility level $P_\textrm{C.L.}$ from a Bayesian posterior distribution is the shortest interval satisfying 
\begin{equation}
    \int_{\theta_L}^{\theta_U} p(\theta\,|\,\mathcal{D})\,\mathrm{d}\theta = P_\textrm{C.L.} 
\end{equation}
where $p(\theta\,|\,\mathcal{D})$ denotes the marginalized posterior density for parameter $\theta$ based on data $\mathcal{D}$.

\begin{figure}[htbp]
  \centering
  \includegraphics[width=\textwidth]{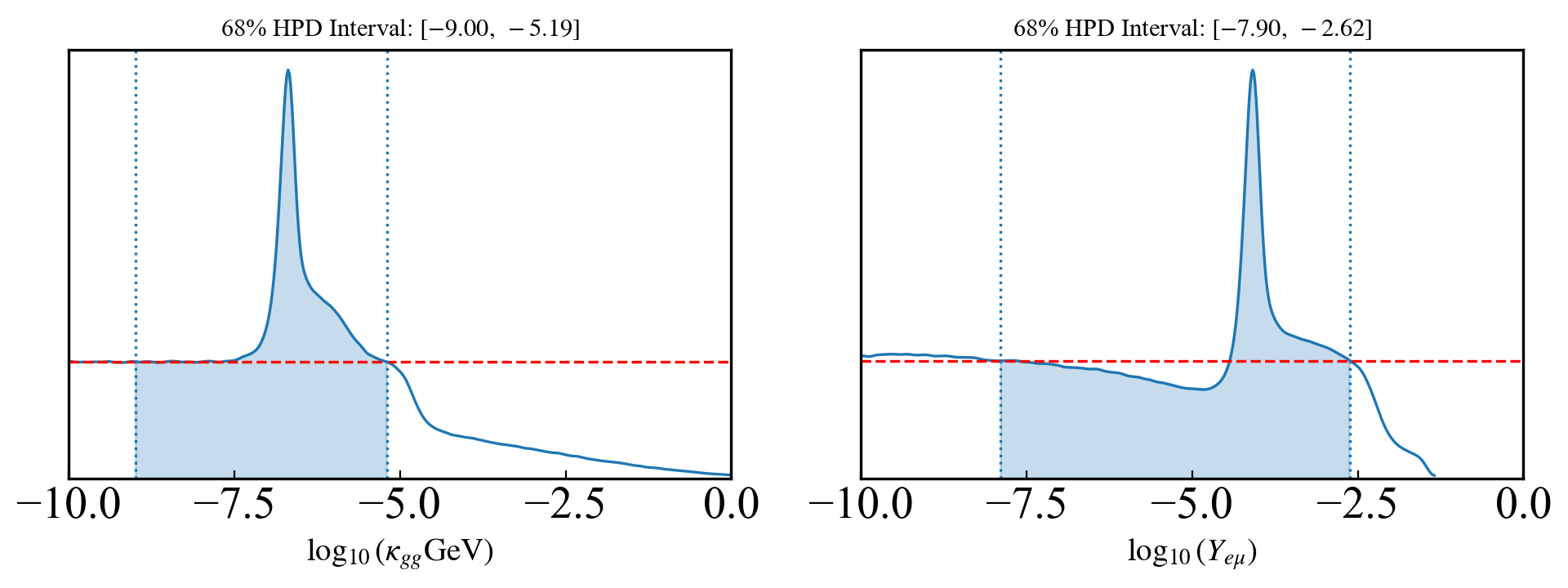}
  \caption{Unimodal HPD intervals for $\log_{10} (\kappa_{gg}\textrm{GeV})$ (left) and $\log_{10}(Y_{e\mu})$ (right). The vertical (blue) dotted lines represent $\theta_L$ and $\theta_U$, respectively.  And the horizontal (red) dotted lines represent the density value at $p=p(\theta_L\,|\,\mathcal{D}) = p(\theta_U\,|\,\mathcal{D})$.}
  \label{fig:HPD}
\end{figure}

\begin{figure}[htbp]
  \centering
  \includegraphics[width=0.6\textwidth]{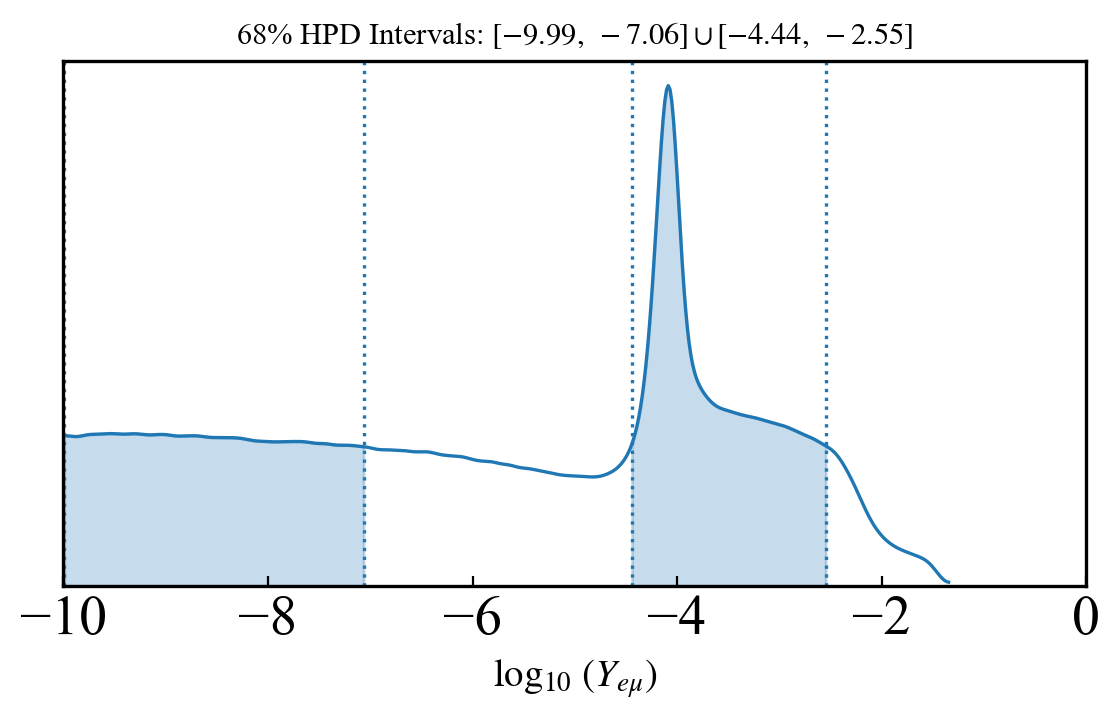}
  \caption{Strict HPD intervals for $\log_{10}(Y_{e\mu})$}
  \label{fig:HPDeu}
\end{figure}

For a unimodal posterior, this requires that \cite{Chen01031999} every point inside the HPD interval has higher posterior density than every point outside it, and the densities at the two endpoints are equal, $p(\theta_L\,|\,\mathcal{D}) = p(\theta_U\,|\,\mathcal{D})$, as shown in Fig.\ref{fig:HPD} (left panel) for $\log_{10}(\kappa_{gg}\textrm{GeV})$. Here we use the \texttt{hdi} function from python library of \textsc{ArviZ}~\cite{kumar2019arviz} to obtain the HPD intervals.

Yet for posteriors with more complicated structure, the HPD region shouldn't be a single connected interval. The relevant case is our $\log_{10}(Y_{e\mu})$ posterior.  When forcing the credible interval to be unimodal and thus connected, as shown in Fig.\ref{fig:HPD} (right panel), the posterior features a broad plateau at small values and a sharp peak near the maximum-likelihood peak, separated by a shallow valley where the marginalized posterior density drops below the inclusion threshold. Strictly, the $68\%$ HPD region for $\log_{10}(Y_{e\mu})$ should therefore split into two regions~\cite{Chen01031999}: the plateau and the peak constitute separate high-density regions, with the valley excluded. In practice, this could be implemented by the \texttt{multimodal=True} option of the \textsc{ArviZ} \texttt{hdi} function. Results are shown in Fig.~\ref{fig:HPDeu}, where the two disjoint intervals correspond to two physically distinct constraints: a vanishing $Y_{e\mu}$ preferred by the low-energy cLFV constraints, and a nonzero value driven by the CMS excess near the peak.

It should be noted that the pattern of two disjoint credible regions of $Y_{e\mu}$ might also be expected to appear in a similar way for $\kappa_{gg}$. But a considerably lower peak value of $\kappa_{gg}$ makes it impossible to resolve into two disjoint preferred regions.

\section{Contribution to $Z$ Boson LFV Di-lepton Decay}
\label{app:Z}
Although not included in the MCMC analysis, the couplings $\phi \bar{\ell}_\alpha\ell_\beta$ can also induce the cLFV di-lepton decay $Z\to \bar{\ell}_\alpha\ell_\beta$ ($\alpha \neq \beta$) at one loop, through the diagrams shown in Fig.~\ref{fig:Zllphi}:

\begin{figure}[htbp]
  \centering

  \begin{tikzpicture}[baseline=(initial.base)]
    \begin{feynman}
      \vertex (initial) {$Z$};
      \vertex [right=1.5cm of initial] (Ztt); 
      \vertex [above right=1.2cm of Ztt] (Yta); 
      \vertex [right=1.2cm of Yta] (a) {$\ell_{\alpha}$};
      \vertex [below right=1.2cm of Ztt] (Ytb);
      \vertex [right=1.2cm of Ytb] (b) {$\bar\ell_{\beta}$};

      \diagram*{ (initial) -- [photon] (Ztt), 
        (Ztt) -- [fermion, edge label=$\tau$] (Yta) -- [fermion] (a), 
        (Ztt) -- [anti fermion, edge label'=$\bar\tau$] (Ytb) -- [anti fermion] (b),
        (Yta) -- [scalar, edge label=$\phi$] (Ytb)};

      \fill[black] (Yta) circle (2pt); 
      \fill[black] (Ytb) circle (2pt); 
      \node[above=0.2cm of Yta] {$Y_{\ell_\alpha \tau}$}; 
      \node[below=0.2cm of Ytb] {$Y_{\ell_\beta \tau}$};
    \end{feynman}
  \end{tikzpicture}
  \quad
  \begin{tikzpicture}[baseline=(initial.base)]
    \begin{feynman}
      \vertex (initial) {$Z$};
      \vertex [right=1.5cm of initial] (Ztt); 
      \vertex [above right=1.5cm of Ztt] (a) {$\ell_{\alpha}$};
      \vertex [below right=0.75cm of Ztt] (Yta);
      \vertex [below right=1cm of Yta] (Ytb);
      \vertex [below right=0.75cm of Ytb] (b) {$\bar\ell_{\beta}$};

      \diagram*{ (initial) -- [photon] (Ztt) -- [fermion] (a),
      (Ztt) -- [anti fermion, edge label'=$\bar\ell_\alpha$] (Yta)
            -- [anti fermion, edge label=$\bar\tau$] (Ytb)
            -- [anti fermion] (b),
      (Yta) -- [scalar, bend right=60, edge label'=$\phi$] (Ytb)};

      \fill[black] (Yta) circle (2pt); 
      \fill[black] (Ytb) circle (2pt); 
      \node[above right=0.05cm of Yta] {$Y_{\ell_\alpha \tau}$}; 
      \node[above right=0.05cm of Ytb] {$Y_{\ell_\beta \tau}$};
    \end{feynman}
  \end{tikzpicture}
  \quad
  \begin{tikzpicture}[baseline=(initial.base)]
    \begin{feynman}
      \vertex (initial) {$Z$};
      \vertex [right=1.5cm of initial] (Ztt); 
      \vertex [below right=1.5cm of Ztt] (b) {$\bar\ell_{\beta}$};
      \vertex [above right=0.75cm of Ztt] (Ytb);
      \vertex [above right=1cm of Ytb] (Yta);
      \vertex [above right=0.75cm of Yta] (a) {$\ell_{\alpha}$};

      \diagram*{ (initial) -- [photon] (Ztt) -- [anti fermion] (b),
      (Ztt) -- [fermion, edge label'=$\ell_\beta$] (Ytb)
            -- [fermion, edge label=$\tau$] (Yta)
            -- [fermion] (a),
      (Ytb) -- [scalar, bend right=60, edge label'=$\phi$] (Yta)};

      \fill[black] (Yta) circle (2pt); 
      \fill[black] (Ytb) circle (2pt); 
      \node[above=0.2cm of Yta] {$Y_{\ell_\alpha \tau}$}; 
      \node[left=0.1cm of Ytb] {$Y_{\ell_\beta \tau}$};
    \end{feynman}
  \end{tikzpicture}
  \caption{Diagrams for $Z \to \ell_{\alpha} \bar\ell_{\beta}$ via scalar $\phi$
  exchange.}
  \label{fig:Zllphi}
\end{figure}
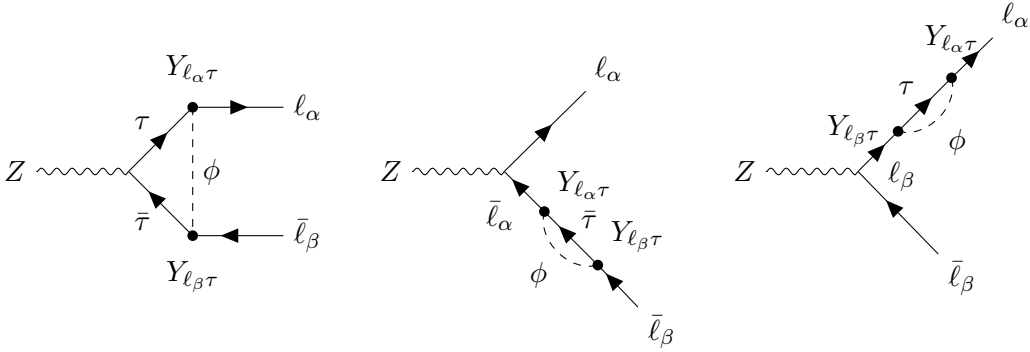
 
One can obtain the branching ratio using the toy model in this work and following the calculation in Ref.~\cite{Goto:2015iha},
\begin{equation}
  \textrm{Br}_{Z\to \ell_{\alpha} \ell_{\beta}}^\textrm{th} = 7.44297\times10^{-8} Y_{\ell_{\alpha}\tau }^2 Y_{\ell_{\beta}\tau}^2,\quad \alpha \neq \beta.
\end{equation}

The most stringent constraints are given by Ref.~\cite{CMS:2025wqy} and Ref.~\cite{ATLAS:2021bdj} at 95\% confidence level,
\begin{align}
  \textrm{Br}_{Z\to e\mu}^\textrm{lim} = 1.9\times10^{-7},\\
  \textrm{Br}_{Z\to e\tau}^\textrm{lim} = 5.0\times10^{-6},\\
  \textrm{Br}_{Z\to \mu\tau}^\textrm{lim} = 6.5\times10^{-6}.
\end{align}
The resulting constraints on the couplings are:
\begin{align}
  \log_{10}{Y_{e\tau}} + \log_{10}{Y_{\mu\tau}} < 0.2035,\\
  \log_{10}{Y_{e\tau}} + \log_{10}{Y_{\tau\tau}} < 0.9136,\\
  \log_{10}{Y_{\mu\tau}} + \log_{10}{Y_{\tau\tau}} < 0.9706,
\end{align}
of which the constraint lines are far higher than the 95\% credible region in Fig.~\ref{fig:res_eu}, Fig.~\ref{fig:res_et} and Fig.~\ref{fig:res_ut}. These constraints therefore have negligible impact on the Bayesian analysis and the conclusions of this work.

\bibliography{bib/limits, bib/future, bib/pheno, bib/tools} 
\end{document}